\documentclass[11pt]{article}
\usepackage{amsmath}
\usepackage{graphicx}
\usepackage{dcolumn}
\usepackage{bm}
\input{tcilatex}

\begin{document}

\title{Nuclear spintronics:\\
quantum Hall and nano-systems}
\author{\textnormal{Israel D. Vagner} \\
\\
{\small {\textit{RCQCE -}} }\\
{\small {\textit{Research Center for Quantum Communication Engineering,}} }\\
{\small {\textit{at Department of Communication Engineering,}} }\\
{\small {\textit{Holon Academic Institute of Technology,}} }\\
{\small {\textit{52 Golomb St., Holon 58102, Israel}} }\\
{\small {\textit{and}} }\\
{\small {\textit{Grenoble High Magnetic Fields Laboratory, }}}\\
{\small {\textit{Max-Planck-Institute f\"{u}r Festk\"{o}rperforschung and
CNRS, }}}\\
{\small {\textit{25 Avenue des Martyrs, BP166, F-38042, Grenoble, Cedex 9,
France }}}\\
{\small {\textit{e-mail: vagner\_i@hait.ac.il}}}\\
{\small {\textnormal{Received 4 January 2004}}}}
\date{}
\maketitle

\begin{abstract}
The electron spin transport in condensed matter, \textbf{Spintronics}, is a
subject of rapidly growing interest both scientifically and from the point
of view of applications to modern and future electronics. In many cases the
electron spin transport cannot be described adequately without accounting
for the hyperfine interaction between electron and nuclear spins. Under
extreme conditions of high magnetic fileds, ultra-low temperatures, ultra
high isotopical cleanness etc., the nuclear spins in these sytems are very
promising candidates for the qubits: the basic elements of emerging quantum
memory, logics and hopefully quantum computers.

Here we review the progress in the \textbf{Nuclear Spintronics} i.e. in
physics and applications of hyperfine interactions in such exotic systems,
as superconducting, quantum Hall, mesoscopic and nano-systems.\newline
\newline
\newline
\end{abstract}

\newpage

\section{Introduction}

The electron spin transport is playing growing role in the fastly developing
new directions of the modern High-Tech electronics, the so-called \textbf{%
Spintronics} \cite{Prinz99,ALS02Bk}. In most solids the conduction electron
system is coupled by hyperfine interaction \cite{AbragBk61,Zahar84Bk} to the
underlying system of nuclear spins, which may profoundly modify the electron
spin transport at low temperatures, thus leading to \textbf{Nuclear
Spintronics} effects, which will be described in this review.

On the other hand, the application of high magnetic fields is a very
powerful tool for studying the electronic properties of a large variety of
metals, semiconductors and superconductors. Due to the Landau quantization
of electron motion in sufficiently strong magnetic field, most of the
transport properties, such as magnetization, conductivity etc. experience
magnetic quantum oscillations (QO) \cite{ShoenbBk}. The Landau quantization
is most spectacularly manifest in the electronic magneto-transport in low
dimensional conductors. Striking examples are the celebrated quantum Hall
effects (QHE) \cite{KDP80}.

Apart from the anomalous enhancement of the well known QO in two-dimensional
electron systems (2DES) one expects also strong QO in physical properties
which are not sensitive to the magnetic field in isotropic three dimensional
metals. It was suggested in \cite{VME82} that in quasi-two-dimensional
metals under strong magnetic fields the nuclear spin lattice relaxation rate 
$T_{1}^{-1}$ should exhibit strong magnetic oscillations.

This should be compared with the Korringa relaxation law \cite{AbragBk61}
usually observed in three-dimensional normal metals, which results in a
magnetic field independent nuclear spin-relaxation rate. This line of
research seems to be useful in dense quasi-two-dimensional electronic
systems, as is the case of synthetic metals (GIC's etc.) and low-dimensional
organic compounds.

A completely new line of research, the hyperfine interaction between nuclear
and electron spins in low dimensional and correlated electron systems, has
been developed during the two last decades both theoretically [8~-~39], 
and experimentally [40 - 59]. 
The growing attention is attracted by the hyperfine interactions in a)~QHE:
theory 
[8 - 19] and experiment 
[40~-~53]; b)~mesoscopics and nano-systems: theory 
[20 - 26] and experiment 
[54~-~55]; c)~normal and superconducting metals: theory 
[27 - 31] and experiment 
[58 - 59]. Very recently nuclear spin memory devices \cite{NucSpMem} and the
indirect hyperfine interaction\ via electrons between nuclear spin qubits in
semiconductor based quantum computer proposals 
[33~-~39] attracts sharply growing attention.

Here I will outline the main theoretical concepts and some experimental
achievements in the new and quickly developing field of \textbf{Nuclear
Spintronics}.

\section{Quantized nuclear spin relaxation}

\subsection{Korringa law}

In metals and doped semiconductors, usually, the leading contribution to the
spin - lattice relaxation process is due to the \textit{hyperfine Fermi
contact} interaction between the nuclear spins and the conduction electron
spins \cite{AbragBk61}. This interaction is represented by the Hamiltonian: 
\begin{equation}
\hat{H}_{int}=-\gamma _{n}\hbar \vec{I}_{i}\cdot \vec{H}_{e},  \label{Hint1}
\end{equation}
where $\gamma _{n}$ is the nuclear gyromagnetic ratio, $I_{i}$ is the
nuclear spin and $H_{e}$ is the magnetic field on the nuclear site, produced
by electron orbital and spin magnetic moments: 
\begin{equation}
\vec{H}_{e}=-g\beta \sum_{e}{\frac{8\pi }{3}}\hat{s}_{e}\delta \left( \vec{r}%
_{e}-\vec{R}_{i}\right) .  \label{He1}
\end{equation}
Here $\vec{r}_{e}$ is the electron radius-vector, $\hat{s}_{e}$ is the
electron spin operator, $\beta =e\hbar /m_{0}c$ is the Bohr magneton and $g$
is the electronic $g$-factor.

The nuclear spin-lattice relaxation rate $T{_{1}^{-1}}$, caused by the
hyperfine Fermi contact interaction between the nuclear spins and the
conduction electron spins, is related to the local spin-spin correlation
function through the equation: 
\begin{equation}
{T_{1}^{-1}}{}={}{\frac{32\pi ^{2}}{9}}\gamma _{n}^{2}g^{2}\beta ^{2}{{\int }%
_{-{\infty }}^{{\infty }}e^{{-i{\omega }_{n}}{t}}}{\left\{ {<S}{^{+}\left( {%
\mathbf{R},t}\right) S^{-}}{\left( {\mathbf{R},0}\right) >}\right\} dt},
\end{equation}
where $S{^{+}\left( \mathbf{R}\right) }$, ${S^{-}\left( \mathbf{R}\right) }$
are the transverse components of the electron spin density operator at the
nuclear position \textbf{R}, and $\omega _{n}$ is the nuclear magnetic
resonance frequency.

\begin{figure}[tbp]
\begin{center}
\includegraphics[width=3.0485in]{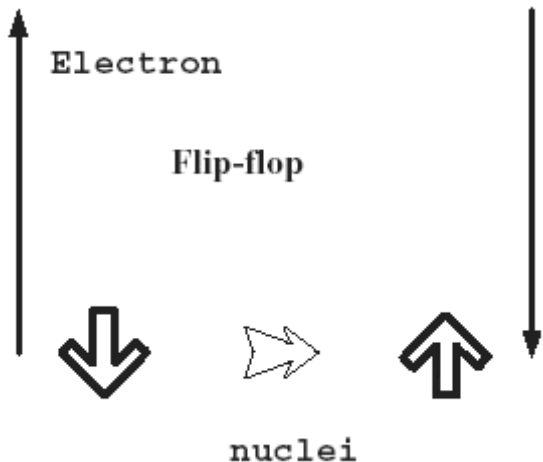} 
\end{center}
\caption{Simultaneous electron and nuclear
spin flips induced by the contact hyperfine interaction.}
\end{figure}

The rate of the nuclear spin-relaxation in metals is, usually, proportional
to the temperature and to the square of the electronic density of states at
the Fermi energy (the Korringa law \cite{AbragBk61}). This follows from the
following expression: 
\begin{equation}
{\frac{1}{T_{1}}}\propto \int_{0}^{\infty }\mid <i\mid V\mid f>\mid ^{2}\rho
(E_{i})\rho (E_{f})f(E_{i})[1-f(E_{f})]\delta (E_{f}-E_{i}+\gamma _{N}H_{0}).
\end{equation}

At low temperatures: $f(E)[1-f(E)]\propto k_{B}T$ $\partial f/\partial E$,
where $k_{B}T$ is the temperature in the energy units, we arrive at the
linear in temperature dependence of \ $1/T_{1}$ 
\begin{equation}
T_{1}^{-1}\sim k_{B}T\rho ^{2}\left( E_{F}\right) ,
\end{equation}
which is the well known Korringa law \cite{AbragBk61}. Here $\rho \left(
E_{F}\right) $ is the electron density of states at the Fermi level.

\subsection{Activation law for $1/T_{1}$ in QHE\ systems.}

In high magnetic fields and in systems with reduced dimensionality this
simple argumentation does not hold, since the electron spectrum acquires
field induced \cite{VM88} or size quantized \cite{VRWZ98,LGAA02} energy gaps.

It was conjectured in \cite{VM88} and confirmed experimentally, see e.g. 
\cite{Berg90,Tycko95}, that in QHE systems the nuclear spin relaxation rate
should have an activation behavior 
\begin{equation}
T_{1}^{-1}\sim \exp \left\{ -\frac{\Delta \left( B\right) }{k_{B}T}\right\} ,
\label{T1exp}
\end{equation}
where $\Delta \left( B\right) $ is either $g\mu _{B}B$, the electron Zeeman
gap (odd filling factors) or $\hbar \omega _{c}$, the Landau levels gap
(even filling factors), instead of the usual Korringa law. The discreteness
of the electron spectrum manifests, at finite temperatures, in an activation
type of the magnetic field dependence of the nuclear spin relaxation rate, $%
T_{1}^{-1}$as it seen from Eq. (\ref{T1exp}). This behaviour is similar to
that of the magnetoresistance $\rho _{xx}$ in the QHE, \cite{VM88}, see
Fig.~2.

\begin{figure}[tbp]
\begin{center}
\includegraphics[width=5.0073in]{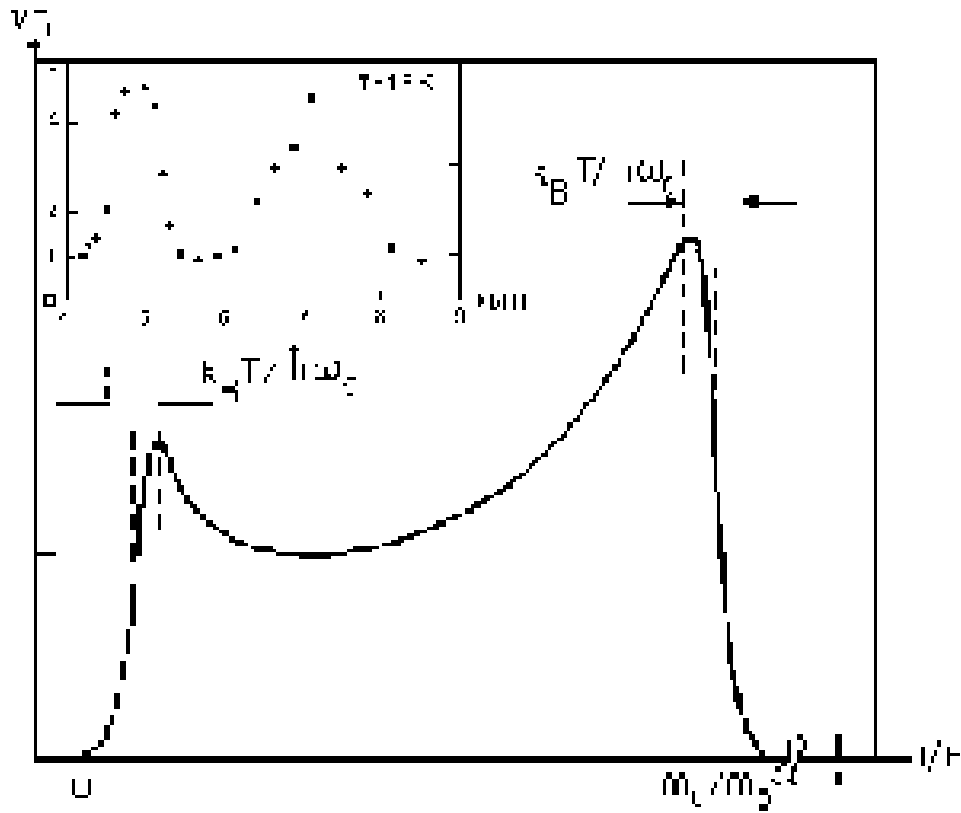} 
\end{center}
\caption{Theoretically predicted 
\protect\cite{VM88}, magnetic field dependence of the nuclear spin
relaxation rate in quantum Hall systems. The inset shows schematically the
experimental data \protect\cite{Berg90}. In the inset the vertical axis is
in msec$^{-1}$.}
\end{figure}

\subsubsection{Energy conservation}

This unusual magnetic field dependence of the nuclear spin relaxation
reflects the fact that the energy gaps in the spectrum of two-dimensional
electrons in strong magnetic fields (either Zeeman splitting or the Landau
levels gap) are three orders of magnitude larger than the nuclear Zeeman
energy. Indeed, the energy needed to reverse the spin of an electron in the
external magnetic field $H_{0}$ is $\Delta E_{el}=2g\mu _{B}H_{0}$, which is
much larger (by a factor of $M_{n}/m_{e}{\simeq }10^{3}$, $M_{n}$ and $m_{e}$
being the nuclear and free electron masses) than the energy $\gamma
_{n}H_{0} $ provided by reversing the nuclear spin. Therefore the delta
function in Eq. (2) can not be realized and the simultaneous spin flip of
the nuclear and the electron spins (flip-flop) in Landau levels, Fig.~3, is
severely restricted by the energy conservation.

\begin{figure}[tbp]
\begin{center}
\includegraphics[width=4.4443in]{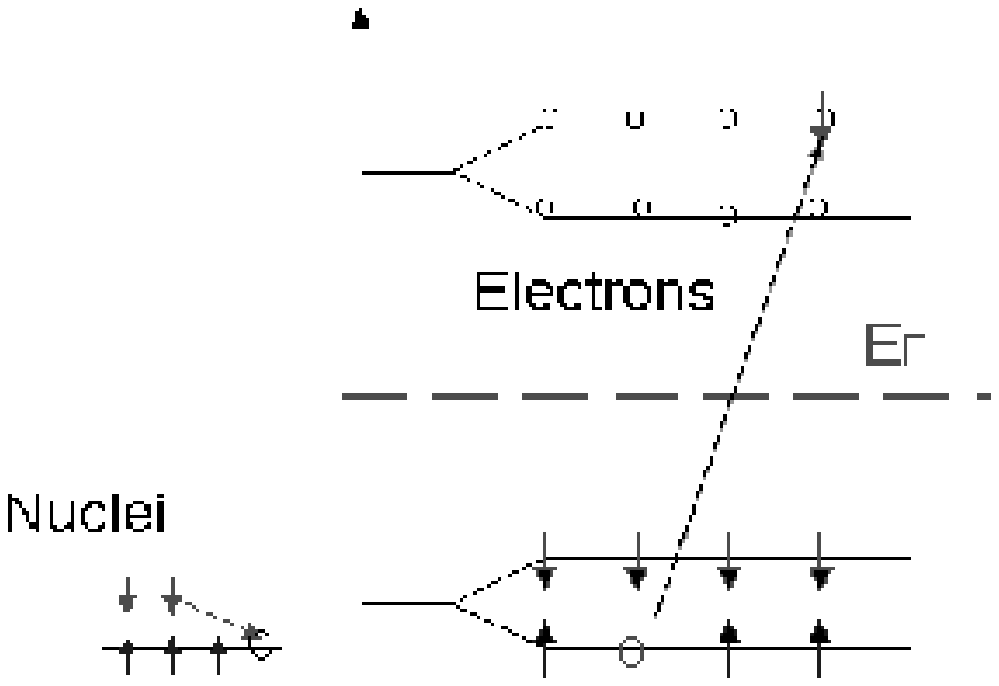} 
\end{center}
\caption{Flip-flop in Landau levels. Even
filling. }
\end{figure}

On the other hand, in isotropic $3D$ electron systems, in a strong magnetic
field : $\hbar \omega _{c}>k_{B}T$, the kinetic energy of the electron
motion parallel to the field should change in order to ensure the energy
conservation of the process shown in Fig.~4. Thus, in the ''isotropic'' $3D$
model, the electron spin - flip will be accompanied by a simultaneous change
of the Landau level and of the kinetic energy parallel to the field, E$_{z}$
according to 
\begin{equation}
\Delta \epsilon _{z}=\hbar \omega _{c}(n^{\prime }-n)+\ \gamma
_{n}H_{o}-\hbar \omega _{z} .  \label{DelEz}
\end{equation}

While this is impossible for an ideal $2D$ system in a strong magnetic
field, it may take place in quasi-two-dimensional conductors, as is the case
in superlattices, for certain regions of parameters.

\begin{figure}[tbp]
\begin{center}
\includegraphics[width=4.5593in]{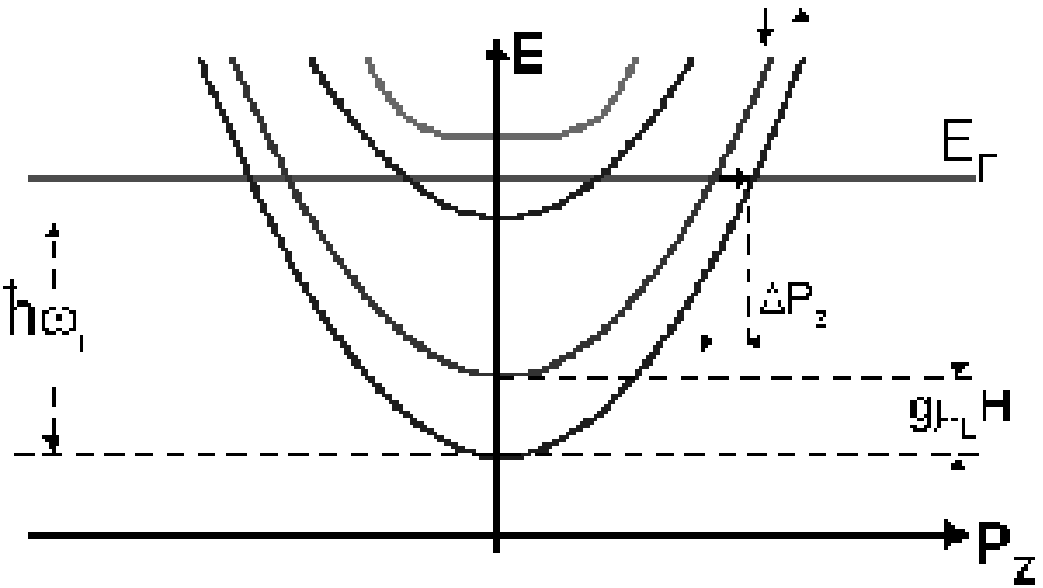} 
\end{center}
\caption{Spin splitted Landau levels in $3D$ electron systems.}
\end{figure}

Because of the existence of energy gaps in the electron spectrum of a 2DES
under strong magnetic fields (the QHE systems), finite nuclear spin
relaxation times $T_{1}$ could be expected only if 2DES is subjected to
different kinds of external potentials.

\subsubsection{Short range impurities}

Broadened Landau levels may overlap, thus providing a nonzero density of
states of both spin projections at the Fermi energy. This will result in
finite relaxation time for excited nuclear spins interacting with the
conduction electrons on the Landau levels. Description of these processes,
performed in the framework of the finite temperature Matzubara diagram
techniques, is presented in \cite{MV90,AMcD91}.

\subsubsection{Edge states}

An important question in the quantum Hall effect theory is the role of the
edge states: electronic orbits magnetically confined to the sample boundary.
While the number of edge states may be small compared to the ''bulk''
states, their contribution to $T_{1}$ can be very important, since they
possess a homogeneous energy spectrum \cite{VMS92}.

In sufficiently clean heterojunctions, however, where the fractional QHE
(FQHE) and Wigner crystallization could be observed, the mechanisms
mentioned above are extremely inefficient.

\subsubsection{Phonon assisted mechanism}

At finite temperatures the energy conservation in the flip-flop process can
be fulfilled by absorbing a phonon [13-15]. 

The mechanism, considered in \cite{KVX94}, consists of two processes in
which, first, electron and nuclear spins are simultaneously reversed by the
hyperfine interaction, and, second, only electron spins are reversed by
their coupling to the lattice strain. In these processes, a phonon is either
absorbed or emitted to satisfy the energy conservation law.

Both the Zeeman split-level energy ($\hbar \omega _{z}$) and chemical
potential ($\mu $) depend on a magnetic field. The dominant contribution to
nuclear relaxation comes from the transition between two spin-split levels
that are near the chemical potential. The contributions from other Landau
levels that are far away from the chemical potential are proportional to $%
exp[-n\hbar \omega _{c}/k_{B}T]$ and are negligible under the QHE condition.
The value at a minimum depend strongly on the value of exchange interaction.

At low temperatures ($k_{B}T\ll \hbar \omega _{z}$), the nuclear spin
relaxation rate varies exponentially with inverse temperature (i.e., $%
T_{1}^{-1}\propto exp[-\hbar \omega _{z}/k_{B}T]$) because the number of
thermally excited phonons in the reservoir becomes exponentially small as
the temperature is decreased. At higher temperatures ($\hbar \omega _{z}\ll
k_{B}T\ll \hbar \omega _{c}$), on the other hand, the relaxation rate varies
linearly with temperature (i.e., $T_{1}^{-1}\propto k_{B}T/\hbar \omega _{z}$%
).

Nucleus-mediated spin-flip transitions in GaAs quantum dots have been
studied in \cite{Erlingson01}. In these papers the electron spin relaxation
times were calculated taking into account the mechanism, where the phonons
provide the necessary energy and nuclear spins take care for the spin
conservation during the electron transition between Zeeman split states in
quantum dots.

\subsubsection{Dipole - dipole interaction}

The nuclear spin relaxation in 2DEG, associated with the dipole-dipole
interaction between the nuclear and the electronic spins was studied by
Ovchinnikov et al.~\cite{OVD94}. The principle physical difference between
these two mechanisms is that while in the contact interaction the total spin
should be conserved and an electron has to flip its spin in order to relax
the nuclear spin, the conserved quantity in the dipole-dipole interaction is
the total (spin plus orbital) momentum. Therefore an electron can relax a
nuclear spin without changing its own spin state, just by shifting its
center of orbit. This mechanism should be prevailing in clean samples, i.e.
under the FQHE conditions.

\section{Electron interaction}

\subsection{Spin-excitons}

Much of the recent attention paid to hyperfine interactions under conditions
of the quantum Hall effect is connected with correlation effects in 2DES.
This is based on the notion of a spin-exciton: the elementary excitation
over the Zeeman gap dressed by the Coulomb interaction \cite{BIE81,KH84}%
.This results in a strong enhancement (up to a factor of 100, as is the case
in GaAs) of the effective $g(k)$ -factor.

\begin{figure}[tbp]
\begin{center}
\includegraphics[width=4.6216in]{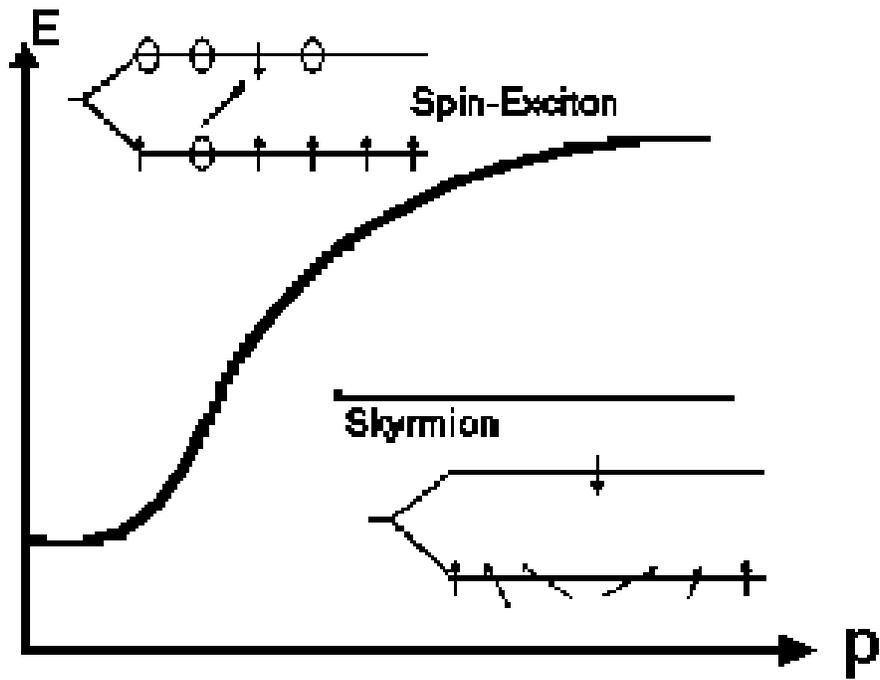} 
\end{center}
\caption{$2D$ Spin exciton and skyrmion
dispersion: the energy gap of large momentum excitons is two times larger
than the skyrmion's one. In the insets the $2D$ electron spin arrangements
for spin excitons (left) and skyrmions (right) is shown.}
\end{figure}

Due to the Coulomb interaction the spin-excitons are bound states of
electron-hole pairs which, unlike the individual electrons or holes, can
propagate freely under the influence of a magnetic field due to their zero
electric charge. These elementary excitations are, therefore, chargeless
particles with a nearly parabolic dispersion in the low-k limit . At $k=0$
the gap is equal to the ''bare'' Zeeman splitting.

The energy spectrum of spin-excitons on the ground Landau level: n=0 is \cite
{BIE81,KH84}: 
\begin{equation}
E_{ex}^{sp}(k)=\mid g\mid \mu _{B}H_{0}+\sqrt{\frac{\pi }{2}}[1-I_{0}({\frac{%
k^{2}}{4}})exp{{\frac{-k^{2}}{4}}}]{\frac{e^{2}}{\kappa a_{H}}} .
\label{Espex0}
\end{equation}

In the parabolic approximation (small exciton momenta), the dispersion
relation reads: 
\begin{equation}
E_{ex}^{sp}(k)\approx \mid g\mid \mu H_{0}+{\frac{k^{2}}{2m_{se}}} ,
\label{Espex1}
\end{equation}
where
\begin{equation*}
{\frac{1}{2m_{se}}}\equiv {\frac{1}{4}}\sqrt{\frac{\pi }{2}}{\frac{e^{2}a_{H}%
}{\chi \hbar ^{2}}}
\end{equation*}
is the definition of the spin-exciton mass.

The invariance of the energy gap with respect to the electron- electron
interaction is associated with the fact that in creating a
quasielectron-quasihole pair excitation at the same point in space (i.e.
with center of mass momentum $k=0$) the energy decrease due to the Coulomb
attraction is exactly cancelled by the increase in the exchange energy. Thus
the energy gap for the creation of a widely separated (i.e. with $%
k\rightarrow \infty $ ) quasielectron-quasihole pair (large spin-exciton) is
equal to the exchange energy associated with the hole.

External potentials, like long range potential fluctuations, periodically
modulated heterostructures etc., may reduce substantiality the spin-exciton
gap.

\subsubsection{Periodically modulated 2DES}

In \cite{BMVW94} the spin-exciton in a periodically modulated
two-dimensional electron gas under strong magnetic fields was investigated.
It was shown there that a periodic external potential, with a period in the
submicron range, can remove the energy gap in the spin-exciton dispersion at
certain values of the spin-exciton momenta. This may result in several
interesting phenomena. For example, the rate of nuclear spin relaxation and
depolarization processes in heterojunctions will be strongly enhanced.

It is easy to see why the Zeeman energy gap can be removed by the modulation
potential in an ideal model of noninteracting electrons: consider a
noninteracting electron-hole pair with opposite spins,which are subject to
an external magnetic field and a periodic potential with amplitude $V_{0}$
and period $a.$ By appropriately selecting the spatial separation between
the electron and the hole along the modulation direction the energy
difference between them can be reduced to zero. The minimal amplitude $V_{0}$
required to satisfy this condition is $\varepsilon _{sp}/2$, the spatial
separation between the electron and the hole involved is half of the
modulation period $a$,and the corresponding wave number (along the
y-direction) is $k_{y}=a/2l_{B}^{2}$. The picture is significantly more
complicated,however,in the presence of the strong Coulomb interaction, where
bound spin-excitons exist.

\subsection{Quantum vacuum fluctuations}

Maniv et al. \cite{MBVW01}, have considered the effect of vacuum quantum
fluctuations in the QH ferromagnetic state on the decoherence of nuclear
spins. It was shown there that the virtual excitations of spin excitons,
which have a large energy gap (on the scale of the nuclear Zeeman energy)
above the ferromagnetic ground state energy, lead to fast incomplete
decoherence in the nuclear spin system. It is found that a system of many
nuclear spins, coupled to the electronic spins in the $2D$ electron gas
through the Fermi contact hyperfine interaction, partially loses its phase
coherence during the short (electronic) time $\hbar /\varepsilon _{sp}$, \
even under the ideal conditions of the QHE, where both $T_{1}$, and $T_{2}$
are infinitely long. The effect arises as a result of vacuum quantum
fluctuations associated with virtual excitations of spin waves (or spin
excitons ) by the nuclear spins. The incompleteness of the resulting
decoherence is due to the large energy gap of these excitations whereas the
extreme weakness of the hyperfine interaction with the $2D$ electron gas
guarantees that the loss of coherence of a single nuclear spin is extremely
small.

The manipulation of the nuclear spins is carried out through spin flip-flop
processes, associated with the 'transverse' part of the interaction
Hamiltonian $\widehat{H}_{en}$, i.e. $A\sum_{j}\left[ \widehat{I}_{j,+}%
\widehat{S}_{-}\left( \mathbf{r}_{j}\right) +\widehat{I}_{j,-}\widehat{S}%
_{+}\left( \mathbf{r}_{j}\right) \right] $, where
\begin{equation*}
\widehat{I}_{j,+}=\widehat{I}_{j,x}+i\widehat{I}_{j,y},\widehat{I}_{j,-}=%
\widehat{I}_{j,x}-i\widehat{I}_{j,y}
\end{equation*}
\ are the transverse components of the nuclear spin operators, and 
\begin{equation*}
\widehat{S}_{+}\left( \mathbf{r}\right) =\widehat{\psi }_{\downarrow
}^{\dagger }\left( \mathbf{r}\right) \widehat{\psi }_{\uparrow }\left( 
\mathbf{r}\right) ,\widehat{S}_{-}\left( \mathbf{r}\right) =\widehat{\psi }%
_{\uparrow }^{\dagger }\left( \mathbf{r}\right) \widehat{\psi }_{\downarrow
}\left( \mathbf{r}\right)
\end{equation*}
are the corresponding components of the electron spin density operators.
Here $\widehat{\psi }_{\sigma }\left( \mathbf{r}\right) $, $\widehat{\psi }%
_{\sigma }^{\dagger }\left( \mathbf{r}\right) $ are the electron field
operators with spin projections $\sigma =\uparrow ,\downarrow $ . The
'longitudinal' part of $\widehat{H}_{en}$, $A\sum_{j}\widehat{I}_{j,z}%
\widehat{S}_{z}\left( \mathbf{r}_{j}\right) $, which commutes with the
Hamiltonian $\widehat{H}_{0}$, and so leaves the nuclear spin projections
along $\mathbf{B}_{0}$ unchanged, can still erode quantum coherence in the
nuclear spin system \cite{Palma96}.

To simplify the analysis it is assumed that the nuclei under study have spin 
$1/2$, and so the corresponding spin operators are expressed in terms of
Fermionic creation and annihilation operators, $\widehat{c}_{j,\sigma }$, as 
\begin{equation*}
\widehat{I}_{j,+}=\widehat{c}_{j,\uparrow }^{\dagger }\widehat{c}%
_{j,\downarrow },\widehat{I}_{j,-}=\widehat{c}_{j,\downarrow }^{\dagger }%
\widehat{c}_{j,\uparrow },\text{ and }\widehat{I}_{j,z}=\frac{1}{2}\left( 
\widehat{c}_{j,\uparrow }^{\dagger }\widehat{c}_{j,\uparrow }-\widehat{c}%
_{j,\downarrow }^{\dagger }\widehat{c}_{j,\downarrow }\right) .
\end{equation*}
In this case the transverse components $\widehat{I}_{j,+}$, $\widehat{I}%
_{j,-}$, are up to a proportionality constant, just the off diagonal
elements (or coherences) of the density matrix of a single nuclear spin
(qubit) \cite{Cohentan, AbragBk61}.

The decay of these elements with time, which determines the rate of
decoherence of a single qubit, can be thus found from the equations of
motion for the operators $\widehat{c}_{j,\sigma }$ in the Heisenberg
representation 
\begin{equation*}
\widehat{c}_{j,\sigma }\left( t\right) =e^{i\widehat{H}t/\hbar }\widehat{c}%
_{j,\sigma }e^{-i\widehat{H}t/\hbar } .
\end{equation*}

Considering a single nuclear spin and evaluating its rate of decoherence due
to the coupling with a 'bath' of spin excitons, assuming that initially, at
time $t=0$, the electronic system is in its ground (QH ferromagnetic) state $%
|0\rangle $, and neglecting the effect of the nuclear spins on the
electronic (bath) states, and averaging over the 'bath' states, one finds to
lowest order in the hyperfine interaction parameter $\alpha $ that the time
dependence of the coherence $I_{+}$ is given by $I_{+}\left( t\right)
=I_{+}\left( 0\right) J\left( t\right) $, where $J\left( t\right) =\exp
[i\Omega \left( t\right) -\Gamma \left( t\right) ].$ This result is similar
to the expression found by Palma et al. \cite{Palma96} in an artificial
model of pure decoherence, i.e. when energy transfer between the qubit and
its environment is not allowed.

The remarkable feature of this expression is due to the presence of the
energy gap $\varepsilon _{sp}$ in the spin exciton spectrum, which is
typically much larger than the nuclear Zeeman energy $\hbar \omega _{n}$ .
During a short time scale, of the order of $\hbar /\varepsilon _{sp}$, the
coherence $I_{+}\left( t\right) $ of a single nuclear spin diminishes and
then saturates for a very long time (i.e. of the order of the relaxation
time $T_{2}$ ) at $I_{+}\left( 0\right) e^{-\eta \widetilde{C}^{2}}$, where 
\begin{equation*}
\eta =\int_{0}^{\infty }\frac{\widetilde{k}d\widetilde{k}e^{-\frac{1}{2}%
\widetilde{k}^{2}}}{\left[ \widetilde{E}_{ex}\left( k\right) \right] ^{2}}%
\sim 1.
\end{equation*}
For $GaAs/Al_{x}Ga_{1-x}As$ heterostructure the coupling constant $%
\widetilde{C}$ is typically of the order of $10^{-4}$ \ \cite{BMV95b}.

\subsection{Long-range random potential}

\subsubsection{Nuclear spin relaxation rate T$_{1}$}

Iordanskii et al. \cite{IMV91} have studied nuclear spin relaxation taking
into account the creation of spin-excitons \cite{BIE81} in the flip-flop
process. The energy for the creation of a spin-exciton can be provided by
the long range impurity potential in a process, where the electron turns its
spin while its center of orbit is displaced to a region with lower potential
energy.

\begin{figure}[tbp]
\begin{center}
\includegraphics[width=4.0171in]{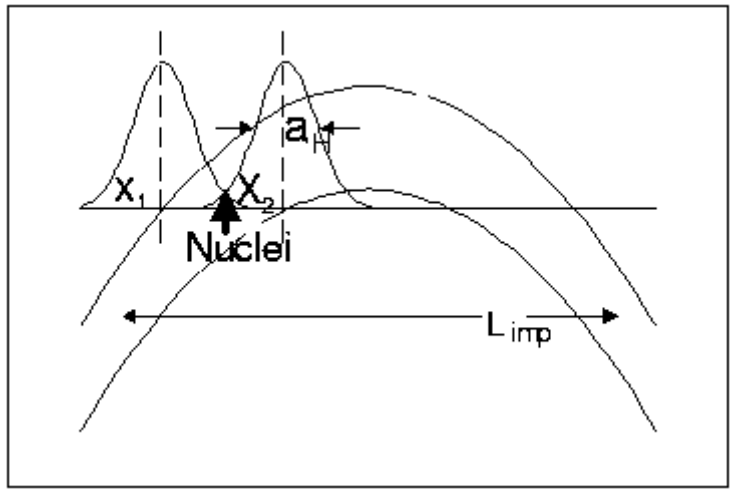} 
\end{center}
\caption{Long-range electrostatic
potential, created by a remote impurity, provides the energy to reverse the
electron spin in the nuclear-electron flip-flop process.}
\end{figure}

As shown in Fig. 6, the overlap of the initial and final location of the
electron wave functions, centered at $x_{1}$ and $x_{2}$ respectively, is: $%
exp[-(x_{1}-x_{0})^{2}/a_{H}-(x_{2}-x_{0})^{2}/a_{H}^{2}]$. Here $x_{0}$ is
the nuclear position. Nuclear spin relaxation by the conduction electron
spin in the vicinity of a potential fluctuation is effective when the
nuclear spin is positioned in the region of the overlapping initial and
final states of the electron wave function.

The energy conservation in the spin-exciton creation process can be written
in the form: 
\begin{equation*}
\mu _{B}gH_{0}+E(p)=x\nabla U .
\end{equation*}
This expression defines the gradient of electric potential caused by the
impurity, sufficient to create a spin-exciton during a flip-flop process.
The probability of finding such a fluctuation is exponentially small: $%
exp[-(\nabla U)^{2}/2<\nabla U^{2}>].$

The momentum \textbf{p} of the spin-exciton is small and therefore the
expansion in p can be performed everywhere in the final expression for
electronic density of states (DOS): 
\begin{equation}
\text{DOS}={\int }ImG(E,x,x)dx=e^{-S}.
\end{equation}
Here $S$ depends on the following combination of physical parameters: 
\begin{equation}
S{\propto }|g|{\frac{({\mu }HE_{c})}{R"(0)}}{\frac{L^{2}}{a_{H}^{2}}},
\end{equation}
$R"(r)\equiv d^{2}R(r)/dr^{2}$ reaches the maximum, usually, at $r=0$.

A complete expression for the nuclear depolarization rate from the golden
rule formula at T=0 \cite{IMV91} is given by 
\begin{equation*}
T_{1}^{1}{\propto }{\nu }_{{\uparrow }}\left( 1-{\nu }_{{\downarrow }%
}\right) {\frac{1}{2\pi ^{2}}}{\int }ImG\left( k,{\omega }\right) {\delta }%
\left( {\omega }-{\omega }_{N}\right) L_{n}\left( k^{2}\right) e^{\frac{%
-k^{2}}{2}}d^{2}kd{\omega },
\end{equation*}
where $\nu _{\uparrow },\nu _{\downarrow }$ are filling factors for the
electrons with spin up and spin down, respectively, n is the highest
occupied Landau level and $L_{n}(k^{2}/2)$ are the Laguerre polynomials.
Here the system of units is used, where $a_{H}^{2}=c\hbar /eH=1$ and ${\hbar 
}=1$.

\subsubsection{Decoherence due to impurities, T$_{2}$}

The dynamics of the nuclear spins is governed by their interactions with
each other and with their environment. In the regime of interest, these
interactions are mediated by 2DES. Various time scales are associated with
this dynamics. The relaxation time $T_{1}$ is related to energy exchange and
thermalization of spins. Quantum mechanical decoherence/dephasing will occur
on the time scale $T_{2}$.

Theoretical calculation\/ of the nuclear-spin dephasing/decoherence time
scale $T_{2}$ for in QHE systems with impurities was presented in \cite
{MPV01}. The hyperfine interaction between the electron and the nuclear
spins, $H_{ne}$, can be split into two parts 
\begin{equation}
H_{ne}=H_{\mathrm{diag}}+H_{\mathrm{offdiag}},  \label{HneMoz}
\end{equation}
where $H_{\mathrm{diag}}$ corresponds to the coupling of the electrons to
the diagonal part of nuclear spin operator $\mathbf{I}_{n}$, and $H_{\mathrm{%
offdiag}}$ --- to its off-diagonal part. It follows \cite{MPV01}, that $H_{%
\mathrm{diag}}$ can be incorporated into the nuclear-spin energy splitting,
redefining the Hamiltonian of the nuclear spin as $H_{n}={\frac{1}{2}}\Gamma
\sigma _{z}$, where $\Gamma =\gamma _{n}\left( B+B_{\mathrm{Knight}}\right) $%
.

The relevant terms in the full Hamiltonian can be expressed solely in terms
of the nuclear-spin operators and spin-excitation operators \cite{MPV01}. It
is assumed there that initially, at time $t=0$, the nuclear spin is
polarized, while the excitons are in the ground state, $|\Psi \left(
0\right) \rangle =|-\rangle \otimes |\mathbf{0}\rangle $ \noindent where $%
|-\rangle $ is the polarized-down (excited) state of the nuclear spin and $|%
\mathbf{0}\rangle $ is the ground state of spin-excitons. Since the
Hamiltonian conserves the total number of elementary excitations in the
system, the most general wave function can be written as 
\begin{equation}
|\Psi \left( t\right) \rangle =\alpha \left( t\right) |-\rangle \otimes |%
\mathbf{0}\rangle +\sum_{\mathbf{k}}\beta _{\mathbf{k}}\left( t\right)
|+\rangle \otimes |\mathbf{1_{k}}\rangle  \label{Eq13Moz}
\end{equation}

\noindent with $|+\rangle $ corresponding to the nuclear spin in the ground
state and $|\mathbf{1_{k}}\rangle $ describing the single-exciton state with
the wave vector $\mathbf{k}$.

Equations of motion for the coefficients $\alpha $ and $\beta _{\mathbf{k}},$%
derived from the Schr\"{o}-dinger equation are: 
\begin{equation*}
i\hbar \dot{\alpha}={\frac{1}{2}}\Gamma \alpha +\sum_{\mathbf{k}}g_{\mathbf{k%
}}\beta _{\mathbf{k}}\mathnormal{\ and\ }i\hbar {\dot{\beta}_{\mathbf{k}}}=-{%
\frac{1}{2}}\Gamma \beta _{\mathbf{k}}+E_{\mathbf{k}}\beta _{\mathbf{k}%
}+\sum_{\mathbf{q}}\phi _{\mathbf{k,q}}\beta _{\mathbf{q}}+g_{\mathbf{k}%
}\alpha .
\end{equation*}

The relaxation rate and the added phase shift of the nuclear-spin
excited-state probability amplitude $\alpha (t)$ are given by the real and
imaginary parts of the pole, respectively: 
\begin{equation*}
{\frac{1}{T_{1}}}={\frac{\pi }{\hbar }}\sum_{\mathbf{k}}g_{\mathbf{k}%
}^{2}\delta \left( \Gamma -E_{\mathbf{k}}\right)
\end{equation*}
and 
\begin{equation*}
\Delta \omega =\ \mathit{P}\sum_{\mathbf{k}}{\frac{g_{\mathbf{k}}^{2}}{%
\Gamma -E_{\mathbf{k}}}}
\end{equation*}
\noindent so that $\alpha (t)\propto \exp [-t/T_{1}+i\Delta \omega t]$. It
is obvious that due to the large gap in the spin-exciton spectrum, $\Gamma
\ll \Delta $, the energy conservation in the flip-flop process can not be
satisfied, and so in the absence of interaction with impurities, $%
T_{1}=T_{2}=\infty $.

Interactions with impurities will modify these solution, and, as a
consequence, the energy conservation condition. In particular, if the
impurity potential is strong enough, it can provide additional energy to
spin-excitons, so that their energy can fluctuate on the scale of order $%
\Gamma $ thus making nuclear-spin relaxation possible. This mechanism was
identified in \cite{IMV91,AMcD91}, and it corresponds to large fluctuations
of the impurity potential $U(\mathbf{r})$, which usually occur with a rather
small probability, so $T_{1}$ is very large for such systems.

The perturbative solution does not describe the energy relaxation ($T_{1}$),
but it does yield the phase shift due to the impurity potential. This phase
shift, when averaged over configurations of the impurity potential, produces
a finite dephasing time, $T_{2}$,which can be calculated considering the
reduced density matrix of the nuclear spin, given by 
\begin{equation}
\rho _{n}(t)=\big[\,\mathrm{Tr}_{e}|\Psi \left( t\right) \rangle \langle
\Psi \left( t\right) |\,\big]_{U}.  \label{Eq26Moz}
\end{equation}

\noindent Here the trace is partial, taken over the states of the
spin-excitons, while the outer brackets denote averaging over the impurity
potential. The trace over the spin-excitons can be carried out
straightforwardly because within the leading-order perturbative
approximation used here they remain in the ground state; all excitations are
virtual and contribute only to the phase shift. The diagonal elements of $%
\rho _{n}(t)$ are not influenced by virtual excitations and remain constant.

The off-diagonal elements of $\rho _{n}(t)$ contain the factors $\exp [\pm
i\Delta \omega _{U}t]$. It is the averaging of these quantities over the
white-noise impurity potential $U(\mathbf{r})$ that yields dephasing of the
nuclear spin. This averaging can be done by utilizing the relation $\left[
\exp (i\phi )\right] _{U}=\exp [-{\frac{1}{2}}\left( [\phi ]_{U}\right)
^{2}] $.

\subsection{Nuclear spin diffusion}

Apart from the direct nuclear spin relaxation, important information about
the electron system can be obtained from nuclear spin diffusion processes.
This is the case when the nuclear spins are polarized in a small part of a
sample as it was experimentally observed in \cite{KPW92,Wald94}.

To explain these experimental observations, Bychkov et al. \cite{BMV95b}
have suggested a new mechanism for indirect nuclear spin coupling via the
exchange of spin excitons. The spin diffusion rate from a given nuclear site 
$\vec{R}_{a}$ within the polarized region is proportional to the rate of
transition probability P($\vec{R}_{a}$) for the polarization of the nuclear
spin $\Downarrow $, located at $\vec{R}_{a}$, to be transferred to a nuclear
spin $\Uparrow $, positioned at $\vec{R}_{b}$, outside the polarized region,
via the exchange of virtual spin excitons, Fig.~7. The virtual character of
the spin-excitons, transferring the nuclear spin polarization, removes the
problem of the energy conservation, typical for a single flip-flop process.
Furthermore, the virtual spin-excitons are neutral entities, which can
propagate freely in the presence of a magnetic field. In this model the
electron interactions play a crucial role: the kinetic energy of a spin
exciton is due to the Coulomb attraction between the electron and the hole.
Thus the proposed mechanism yields the possibility of transferring nuclear
spin polarization over a distance much longer than the magnetic length $\ell
_{B}$. The long range nature of this mechanism is of considerable importance
when the size of the region of excited nuclear spins, $L_{ex}$, is much
larger than the magnetic length $\ell _{B}$.

\begin{figure}[tbp]
\begin{center}
\includegraphics[width=3.0692in]{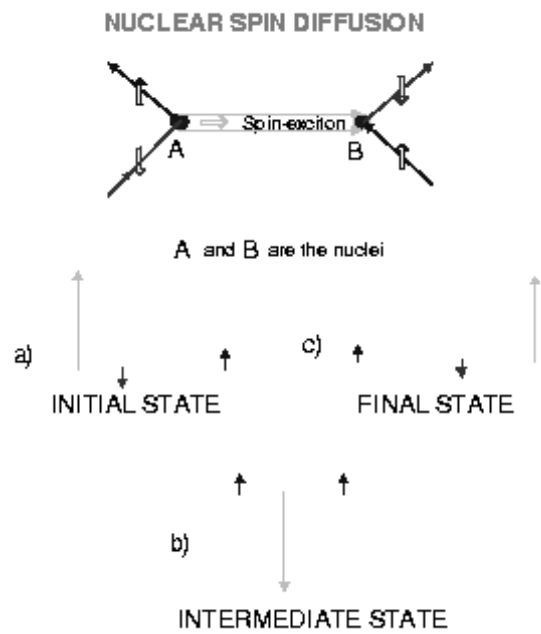} 
\end{center}
\caption{Indirect interaction between two
nuclear spins via conduction electron spin.}
\end{figure}

As it is shown in \cite{BMV95b}, the potential of the nuclear spin - spin
interaction, mediated by the exchange of spin-excitons, is a monotonic
function of the distance between the two nuclei with the asymptotics:
\begin{equation*}
U\left( R_{ab}\right) \propto -\sqrt{\frac{d}{R_{ab}}}e^{-{\frac{R_{ab}}{d}}%
} ,
\end{equation*}
where
\begin{equation*}
d\equiv {\frac{\ell _{B}}{2}}\sqrt{\frac{\epsilon _{c}}{\epsilon _{sp}}} .
\end{equation*}
This is typical for the interaction, mediated by the exchange of
quasiparticles with an energy gap at $q\rightarrow 0$, as is the case for
the spin-exciton dispersion. The range $\Delta R$ of this potential is
determined by the critical wave number 
\begin{equation*}
k_{0}={\frac{2}{\ell _{B}}}\sqrt{\frac{\epsilon _{sp}}{\epsilon _{c}}}
\end{equation*}
as it follows from the uncertainty principle: $\Delta R\cdot k_{0}\simeq 1$.
The negative sign of this interaction corresponds to attraction between the
nuclear spins and may cause, at sufficiently low temperatures, a
ferromagnetically ordered nuclear state in QHE systems.

\subsection{Skyrmions and skyr-nuons}

Spin-excitons constitute the building blocks, from which more complex spin
textures can be formed. For example, at finite densities the spin excitons
\textquotedblright condense\textquotedblright\ \cite{BMV96} into
skyrmion-antiskyrmion pairs. These unusual topological excitations \cite
{lee90,Girvin99} in the spin distribution of real 2DES were observed, in NMR
experiments, near filling factor $\nu =1$ \cite{Barrett94,Tycko95}. Further
evidences for skyrmions in 2DES were found in transport and optical
experiments.

Skyrmions, in QHE systems, are the topologically nontrivial spin excitations
around filling factor $\nu =1$ \cite{lee90} which arise as a condensate of
interacting spin excitons \cite{BMV96}. The Coulomb interaction acts to
enlarge the Skyrmion size while the Zeeman splitting tends to collapse
Skyrmions. The interplay between these factors determines the final
distribution of spins within a Skyrmion, and its characteristic length
scales. The resulting radius $R$ corresponds to the region where both these
energies are of the same value, and grows weakly to infinity as the g-factor
goes to zero \cite{BKMV96}, thus reflecting the importance of the long range
Coulomb repulsion associated with the Skyrmion charge in the zero g-factor
limit.

A canonical $u-v$ transformation from the fully polarized ground state,
where all spins are oriented along a single axis, to a state, consisting of
a macroscopic number of differently oriented spins, each of which is
slightly rotated with respect to its nearest neighbors in space was is
developed in \cite{BMV96}. It is found also that contrary to the lowest
Landau level, these energies are positive for both skyrmions and
antiskyrmions.

It should be noted that the spin-rotation transformation employed in \cite
{BMV96} is unitary and does not change the total number of electrons. Thus
by going to the new state $|\psi >$ from the fully polarized ground state $%
|\psi _{0}>$ the total topological charge does not change. This can be done
if the topological defects are created in pairs of widely separated
skyrmions and the corresponding antiskyrmions with equal and opposite
charges. The total energy of such a skyrmion-antiskyrmion pair, with winding
number $Q=1$, is exactly equal to half the total energy required to create a
well separated electron-hole pair (large spin exciton).

If the Zeeman splitting is not neglected, however, the increase in magnetic
energy associated with the rotated spins determines a length scale for the
spatial size of the defect . To take into account Zeeman spin splitting the
Hartree-Fock (HF) energy functional should be corrected by the additive term 
$-\frac{\epsilon _{sp}}{4\pi }\int d^{2}r(\hat{z}\cdot \vec{n})$ \cite
{BKMV96}, where $\epsilon _{sp}$ is the Zeeman splitting energy. The
corresponding equation for the vectorial field $\vec{n}(\vec{r})$ can be
obtained by variation of the corrected energy functional with respect to $%
\vec{n}$ under the constraint $|\vec{n}|^{2}=1$. The result of such a
calculation is: 
\begin{equation}
\Delta \vec{n}-\vec{n}(\vec{n}\cdot \Delta \vec{n})=l_{sk}^{-2}[(\hat{z}%
\cdot \vec{n})\vec{n}-\hat{z}],  \label{eqforn}
\end{equation}
where
\begin{equation*}
l_{sk}^{-2}\equiv 4\epsilon _{sp}/E(0)l_{H}^{2}=(4/\sqrt{2\pi })|g|(\tilde{a}%
_{B}/l_{H}^{3})
\end{equation*}
and 
\begin{equation*}
\tilde{a}_{B}\equiv \frac{\kappa \hbar ^{2}}{m_{0}e^{2}}
\end{equation*}
is the effective Bohr radius (note that $m_{0}$ is the free electron mass).

\begin{figure}[tbp]
\begin{center}
\includegraphics[width=5.1517in]{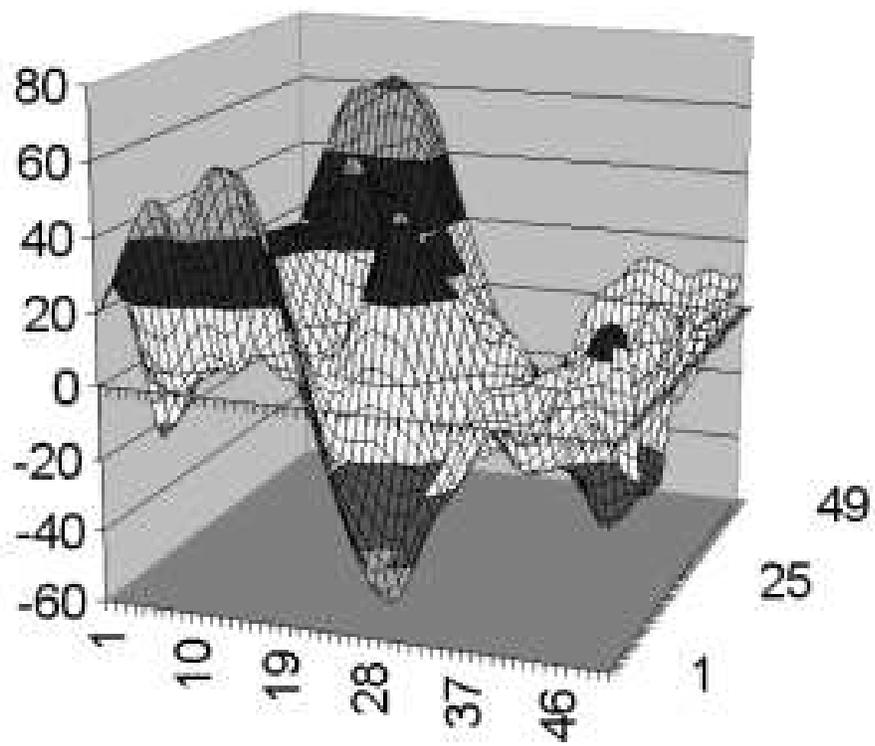} 
\end{center}
\caption{Random hyperfine field potential.}
\end{figure}

In NMR experiments on skyrmions 
[43,46-48,51] the nuclear spins are strongly polarized. The sample
inhomogeneity may result in a strong inhomogeneity of the hyperfine field,
Fig.~8, and therefore spatially varying electron Zeeman splitting.

\begin{figure}[tbp]
\begin{center}
\includegraphics[width=4.2261in]{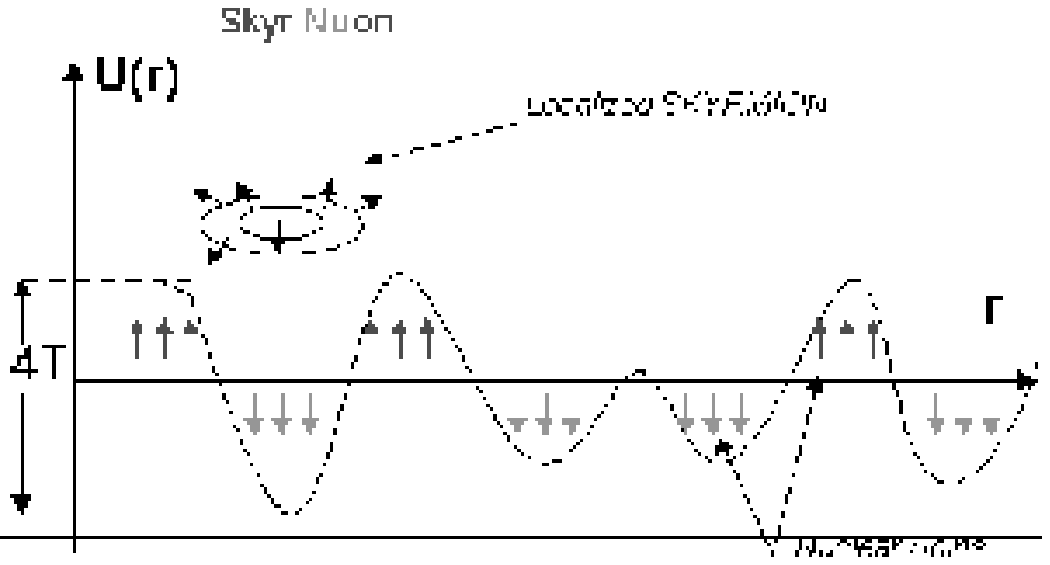} 
\end{center}
\caption{Skyr-nuons, i.e. finite size
skyrmions, localised by a random hyperfine field.}
\end{figure}

This may result in a strong localization of skyrmions \cite{NN99,Barrett01},
resulting in the combined topologically nontrivial electron-nuclear spin
excitation, the skyr-nuon, Fig.~9.

\section{Nonequilibrium nuclear spin polarisation}

As we have seen in previous sections, the nuclear spin polarization, once
created, remains finite for macroscopically long times \cite{Zahar84Bk}.
Intensive experimental studies \cite{Berg90,Wald94,Tycko95} of this
phenomenon have provided a more detailed knowledge on the hyperfine
interaction between the nuclear and electron spins in heterojunctions and
quantum wells. It was observed that the nuclear spin relaxation time is
rather long (up to 10$^{3}\sec $) and the hyperfine field acting on the
charge carriers spins is extremely high, up to 10$^{4}G$ \cite{Berg90,Wald94}%
. The nontrivial physics in this subject is based on a fact that the
discrete nature of the electron spectrum in these systems will result in
exponentially long dependence of the nuclear spin relaxation times $T_{1}$
on the system parameters \cite{VM88}, i.e. T$_{1}\sim $ $\exp \{\Delta /T\}$
(here $\Delta $ is the electron energy level spacing and $T$ is the
temperature). We assume here that similar law should take place also in the
nanostructures with well defined size quantization of the electron spectrum.
Note that in this case $T_{1}$ is very sensitive to the potential
fluctuations, caused by the inhomogeneous distribution of impurities in a
heterojunction \cite{IMV91}.

\subsection{Hyperfine Aharonov-Bohm effect}

A family of new physical effects in nanostructures with strong spin-orbital
coupling appears when the electron spin degeneracy is lifted by a hyperfine
field of polarized nuclei. Indeed, the combined action of a strong nuclear
polarization and the spin-orbit interaction, breaks the time reversal
symmetry in a mesoscopic system. A good illustration for such \textbf{%
meso-nucleo-spinic} effects is the hyperfine Aharonov-Bohm effect (HABE) in
mesoscopic rings \cite{VRWZ98}, caused by hyperfine interaction, as is
explained in what follows.

Persistent currents (PC) in mesoscopic rings reflect the broken clock
wise-anticlock wise symmetry of charge carriers momenta caused, usually, by
the external vector potential. Experimentally PCs are observed when an
adiabatically slow time dependent external magnetic field is applied along
the ring axis. The magnetic field variation results in the oscillatory, with
the magnetic flux quantum $\Phi _{0}=\frac{hc}{e}$ (or its harmonics) period
behavior of the diamagnetic moment (the PC), which is the manifestation of
the usual Aharonov-Bohm effect (ABE).

It was suggested in \cite{VRWZ98} that in a quantum ring with a
nonequilibrium nuclear spin population the persistent current will exist,
even in the absence of external magnetic field. It is shown there, that the
ABE like oscillations of PC with time will appear, during the time interval
of the order of nuclear spin relaxation time $T_{1}$. The physics of this
phenomenon can be understood along the following lines.

The hyperfine field, caused by the nonequilibrium nuclear spin population
breaks the spin symmetry of charged carriers. Combined with a strong spin -
orbital (SO) coupling, in systems without center of inversion \cite{BR84},
it results in the breaking of the rotational symmetry of diamagnetic
currents in a ring. Under the topologically nontrivial spatial nuclear spin
distribution, the hyperfine field produces an adiabatically slow time
variation of the Berry phase of the electron wave function.

The time variation of this topological phase results in observable
oscillations of a diamagnetic moment (the PC). It is one of a series of
\textquotedblright meso-nucleo-spinic\textquotedblright\ effects, which may
take place in mesoscopic systems with broken symmetry, due to the combined
action of the hyperfine field and spin-orbital interaction.

Due to the contact hyperfine interaction Eq. (\ref{Hint1}) it follows that
once the nuclear spins are polarized, i.e. if $\left\langle \sum_{i}\mathbf{I%
}_{i}\right\rangle \neq 0$, the charge carriers spins feel the effective,
hyperfine field $B_{hypf}=B_{hypf}^{o}\exp \left( -t/T_{1}\right) $ which
lifts the spin degeneracy even in the absence of external magnetic field. In
GaAs/AlGaAs one may achieve the spin splitting due to hyperfine field of the
order of the one tenth of the Fermi energy \cite{Berg90,Wald94}.

Let us suppose therefore that the charge carriers spin orientation is
partially polarized during the time interval of the order of $T_{1}$ . It is
quite obvious that the topologically nontrivial spin texture combined with
the spin-orbit interaction will result in a PC.

\begin{figure}[tbp]
\begin{center}
\includegraphics[width=3.6726in]{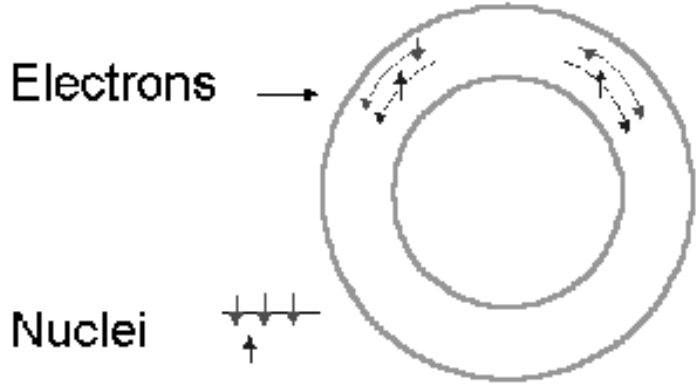} 
\end{center}
\caption{A mesoscopic ring with
inhomogeneously polarised nuclear spins. The nonequilibrium nuclear spin
population will result in Hyper - Aharonov - Bohm effect, as is descriebed
in the text. The left and right rotating electrons with both spin directions
are shown.}
\end{figure}

Taking the Bychkov-Rashba term \cite{BR84} 
\begin{equation}
\widehat{H}_{so}=\frac{\alpha }{\hbar }\sum_{i}\left[ \mathbf{\sigma }%
_{i}\times \mathbf{p}\right] \mathbf{\nu },  \label{HsoRashba1}
\end{equation}
where $\alpha =0.6\cdot 10^{-9}eVcm$ for holes with $m^{\ast }=0.5m_{0}$ ($%
m_{0}$ is the free electron mass), and $\alpha =0.25\cdot 10^{-9}eVcm$ for
electrons, $\mathbf{\sigma }_{i}$,$\mathbf{p}_{i}$ are the charge carrier
spin and momentum and $\mathbf{\nu }$ is the normal to the surface. It can
be rewritten in the form 
\begin{equation}
\widehat{H}_{so}=\mathbf{pA}_{eff},  \label{Hso=pAeff}
\end{equation}
where 
\begin{equation}
A_{eff}^{GaAs}\simeq \frac{\alpha m^{\ast }}{\hbar }\left\langle \sigma
\right\rangle ,  \label{AeffGaAs}
\end{equation}
$\left\langle \sigma \right\rangle $ stands for a nonequilibrium carriers
spin population. Under the conditions of a topologically nontrivial
orientation of $\mathbf{A}_{eff}^{GaAs}$ the wave function of a charge
carrier encircling the ring gains the phase shift similar to the one in an
external magnetic field like in the ordinary ABE. This phase shift can be
estimated as follows 
\begin{equation}
2\pi \Theta =\frac{1}{\hbar }\oint A_{eff}^{GaAs}dl=\frac{m^{\ast }}{\hbar
^{2}}\left\langle \sigma (t)\alpha \right\rangle \sim \frac{m^{\ast }\sigma
(t)\alpha }{\hbar ^{2}}L,  \label{phi1}
\end{equation}
where $L$ is the ring perimeter. To observe the oscillatory persistent
current connected with the adiabatically slow time-dependent $\left\langle
\sigma \left( t\right) \right\rangle ,$ $L$ is supposed to be less than the
phase breaking length. Taking the realistic values for $L\approx 3\mu m$ and 
$\left\langle \sigma \right\rangle \approx 0.05\div 0.1$ we estimate $2\pi
\Theta \sim 5\div 10$ which shows the experimental feasibility of this
effect.

The standard definition of the spontaneous diamagnetic current is 
\begin{equation}
j_{hfso}=-c\frac{\partial F}{\partial \phi }\mid _{\phi _{ext}=0},
\label{jhfso}
\end{equation}
where F is the electron free energy and $\phi _{ext}$ is the external
(probe) magnetic flux. The oscillations of PC arise due to the exponential
time dependence of the phase $\Theta _{eff}^{0}\exp \left\{ -t/T_{1}\right\} 
$ in Eq. (\ref{phi1}), with the time constant $T_{1}.$From the analogy with
the standard ABE in a mesoscopic ring, we expect the following form for a
persistent current in an one dimensional quantum ring at low enough (T $\ll
\Delta $) temperature 
\begin{equation}
j_{hfso}\sim \frac{ev_{F}}{L}\sin (2\pi \Theta _{o}e^{-\frac{t}{T_{1}}}).
\label{jhfso1}
\end{equation}
Here $\Theta _{o}$ is the initial phase value. The marking difference
between the periodical time dependence of standard Aharonov-Bohm
oscillations, which are observed usually under the condition of linear time
variation of the applied magnetic field and the hyperfine driven
oscillations which die off due to the exponential time dependence of the
nuclear polarization.

\subsection{Nuclear spin polarization induced (NSPI) nano-structures}

In \cite{NSPI} it was proposed to use the inhomogeneous hyperfine field to
create so-called nuclear spin polarization induced (NSPI) magnetic
structures, such as magnetic quantum dots, wires, rings, superlattices etc.

The spin splitting ($\mu _{B}B_{hf}$) due to a hyperfine magnetic field can
be comparable to the Fermi energy of 2DEG, so that the electrons in the
region, where nuclear spins are polarized, will occupy the energetically
more favorable states with the spins opposite to $\mathbf{B}_{hf}$. The
inhomogeneous nuclear polarization acts on the electrons as the effective
confining potential $V_{conf}=-\mu _{B}B_{hf}$. This effective confining
potential can be used to create different nanostructures with polarized
electrons in them.

The method of local nuclear spin optical polarization allows to create
different NSPI quantum structures (quantum dots, rings, wires, etc.) using
the same sample and different illumination masks.

The time evolution of the hyperfine magnetic field in NSPI quantum
structures, due to the nuclear spin relaxation and the nuclear spin
diffusion, leads to variation of the number of transverse modes and
corresponding electron energies at a constant gate potential that can be
directly measured by transport experiments.

The dependence of the conductance at 'zero' temperature on the number of
transverse modes in the conductor is given by the Landauer formula $%
G=2e^{2}MT/h$, where $T$ is the average electron transmission probability, $%
M $ is the number of the transverse modes and the factor $2$ stands for the
spin degeneracy. It is assumed that the transition probability $T$ is
independent of the energy in a small interval between the chemical
potentials of the reservoirs. Usually, the number of the transverse modes,
defined by the effective width of the conductor, is controlled by the gate
voltage and the conductance is changed in discrete steps $2e^{2}/h$. It is
underlined in \cite{NSPI} that due to the spin selective effective potential
the height of the conductance steps is just $e^{2}/h$ which is a half of the
conductance quantum $G_{0}=2e^{2}/h$.

The experimental feasibility is based on the method of optical nuclear spin
polarization \cite{Zahar84Bk,Barrett94,Tycko95,Lampel68}. The resolution of
the optical illumination of the sample can be high enough. Usual optic
technique allows to create the light beams of the width of the order of the
wave length ($\sim 500nm$), by using near fields optics the beam width can
be sufficiently reduced ($\sim 100nm$). Hence a NSPI QW of the width of $%
1\mu m$ can be created by the modern experimental technique. In
semiconductor heterostructures having supreme quality, the electron mean
free path can be as large as $100\mu m$ and NSPI QW will operate in the
quantum regime.

\subsection{Hyper-anomalous Hall effect (HAHE)}

In bulk semiconductors, the study of the conduction electron
magnetotransport under\ the influence of the nonequilibrium nuclear spin
polarization was started a long time ago by measurement of the
magnetoresistance in bulk InSb \cite{SKL71}. Very recently a refined
detailed study of the influence of a weak hyperfine field on the Hall effect
in bulk InSb under a strong magnetic field (one occupied Landau level) was
performed \cite{GJ96}.

Apart of the usual Hall effect, caused by the drift of electrons in the
crossed electric and magnetic fields, there exist also another contribution
to the nondiagonal components of the conductivity tensor, the anomalous Hall
effect \cite{NL73,LB-MN73} (AHE), which is caused, in nonmagnetic
semiconductors, by the spin-orbit interaction combined with the carrier
magnetization.

It was proposed in \cite{Bednar98} that in the strong nuclear spin
polarization limit and at low magnetic field this effect may give a leading
contribution to the anomalous Hall effect. Due to this interaction,
electrons with their spin polarization parallel to the magnetization axis
will be deflected at right angles to the directions of the electric current
and of the magnetization while electrons with antiparallel spin polarization
will be deflected in the opposite direction. Thus, if the two spin
populations are not equal there appears a net current in the transverse
direction.

Until now, studies of the anomalous Hall effect have been limited to the
case where the carrier magnetization is induced by the magnetic field. The
magnetic field, however, produces a much larger normal Hall effect which
makes experimental studies quite difficult.

Recording the Hall voltage in a sample with a dynamically polarized nuclear
spins, during a time scale of the order of the nuclear spin relaxation time,
which can be very long on the scale of typical electron equilibration
processes, gives a direct measure of the spin dependent component of the
Hall voltage. If the experiment is designed in such a way that the hyperfine
field is much higher then the external magnetic field the hyperfine field
driven anomalous Hall effect can be observed.

InSb is well suited for observation of the HAHE\ effect because of the
strong spin-orbit coupling and the strong, due to the high atomic numbers of
the atoms involved, hyperfine interaction ($\Delta =0.9$ eV, $E_{g}=0.235$
eV, $g^{\ast }=-51$, $m^{\ast }=0.013m_{0}$). The maximal nuclear-spin
polarization for all isotopes of InSb yields $B_{HF}^{\max }\approx 0.37%
\unit{T}$ \cite{Bednar98}.

It is outlined in \cite{Bednar98} that the HAHE\textit{\ }gives the
possibility to observe genuine spin dependent anomalous Hall effect which
relies on the natural relaxation of nuclear spins and facilitates the
experiment since the anomalous Hall voltage is not hidden by the much larger
ordinary Hall effect.

\subsection{Hyperfine residual resistance and the $\protect\tau _{\protect%
\phi }$ problem}

The possibility that the hyperfine interaction between the conduction
electron spins and nuclear spins may result in hyperfine residual
resistivity (HRR)\textit{\ }in clean conductors at very low temperatures was
studied theoretically by Dyugaev et al. \cite{DVW96}. Apart from the
fundamental nature of this problem, the natural limitations on the mean free
path are decisive in semiconductor based high speed electronic devices, like
heterojunctions and quantum wells. The space periodicity of nuclei plays no
specific role, as long as the nuclear spins are disordered and act as
magnetic impurities \cite{KF94} with the concentration $C_{n}\approx 1$.
This scattering is not operative at extremely low temperatures, in the $\mu
K $ region when the nuclear spins are ferro- or antiferromagnetically
ordered.

The HRR, arising due to the Fermi (contact) hyperfine interaction between
the nuclear and the conduction electron spins can be written in atomic units
as: 
\begin{equation}
V_{en}\approx Z\alpha ^{2}\frac{m_{e}}{M_{n}}Ry  \label{Ven1}
\end{equation}

Here $\hbar =m_{e}=e=1,$ $m_{e},M_{n}$ are the electron and the nucleon
masses, respectively; $Ry=27$ eV and $\alpha =1/137$ is the fine structure
constant.

In metals the effective electron-nuclear interaction constant is \ 
\begin{equation*}
g_{n}\equiv \frac{V_{ne}}{\epsilon _{F}}\approx 10^{-7}Z\frac{Ry}{\epsilon
_{F}}\text{ ,}
\end{equation*}
where the Fermi energy $\epsilon _{F}$ varies in a wide interval $0.01\div 1$%
. $g_{n}$ is $10^{-6}$ for $Li$ and $10^{-3}$ for the rare earth metals.

The total residual resistivity is therefore a sum of the impurity $\rho
_{o}(T\rightarrow 0)\sim C_{o}$ and the nuclear spin $\rho _{n}(T\rightarrow
0)\sim g_{n}^{2}$ contributions: 
\begin{equation}
\rho _{o}^{+}(0^{+})\approx \rho _{oo}(C_{o}+g_{n}^{2}) .
\end{equation}

Here $\rho _{oo}\approx 1$ in atomic units: $\rho _{oo}\approx 10^{-17}$
sec. The nuclear contribution to resistivity starts to be operative when the
impurity concentration is $C_{o\text{ }}\sim g_{n}^{2}$.

In the limit of an ideally pure ($C_{o\text{ }}=0$) metal the universal
residual resistivity $\rho _{URR\text{ }}$is, therefore $\rho _{URR\text{ }%
}\geq \rho _{oo}g_{n}^{2}$ and the mean free path is limited by $%
10^{-8}/g_{n}^{2}$ cm. This yields $10^{4}$ cm in $Li$ and $10^{-2}$ cm for
the rear earth metals. It is interesting to note that in materials with
even-even nuclei (zero spin), like $Ca$, $Ni$, $Fe$, $Ce$ and isotopically
clean graphite $C$, where the electron-nuclear scattering is absent, the 
\textit{HRR} would not be observed.

The temperature and magnetic field dependence of the \textit{HRR} contains
reach information on the hyperfine interaction between the conduction
electrons and the nuclear spins. Usually, the temperature and the magnetic
field dependence of residual resistivity due to nonmagnetic impurities is
due mostly to the mesoscopic effects, and is vanishing in the limit $C_{o%
\text{ }}\rightarrow 0$ . In a magnetic field such that $\mu _{e}H\gg T$ the
magnetic impurities freeze out and the Kondo effect is quenched. In order to
freeze out the nuclear spins however one should apply much higher magnetic
fields, $\mu _{n}H\gg T$ . Therefore in the temperature interval $\mu
_{e}H\gg T\gg $ $\mu _{n}H$ the nuclear spin contribution may prevail even
in metals with magnetic impurities.

In metals like $Li,Na,K,Rb,Cs,Au,Cu$ the nuclear magnetic moments $I\neq 1/2$
and even without external magnetic field their $2I+1$ degeneracy is lifted
partially by the quadrupole effects (in the case of cubic crystal symmetry
the quadrupole splitting of the nuclear levels may happen due to the defects 
\cite{AbragBk61,Rowlend60} and dislocations \cite{Averbuch59}).

While the normal metals have a quite similar electronic structure, the
experimentally observed temperature dependence of the dephasing time $\tau
_{\varphi }$ is quite different. This was shown in \cite{Mohanty99,Esteve1},
where the value of \ $\tau _{\varphi }$ was defined by the magnetoresistance
measurements of long metallic wires $Cu,Au,Ag$ in a wide temperature
interval $10^{-2}<T<10^{^{o}}$K. In $Cu$ and $Au$ wires \cite{Mohanty99}$%
\tau _{\varphi }$ saturates at low temperatures which contradicts the
standard theory \cite{AAK82}. Strangely enough the Ag wires do not show
saturation \cite{Esteve1} at the lowest temperatures, in accordance with 
\cite{AAK82}.

It is conjectured in \cite{DVW00CM} that the influence of the quadrupole
nuclear spin splitting on the phase coherence time $\tau _{\varphi }$ can be
a clue to this puzzle. Indeed, the nuclear spins of \ both $Cu$ and $Au$
have a strong quadrupole moment ($s=3/2$) and may act as inelastic two-level
scatterers once their degeneracy is lifted by the static impurities \cite
{Rowlend60}, dislocations \cite{Averbuch59} and other imperfections. It is
known that (see \cite{KA00} and references therein) the nondegenerate two
level scatterers may introduce inelastic phase breaking scattering of
conduction electrons.

This may be not the case for $Ag$ nuclei since their spin is $s=1/2$. In
this case the quadrupole splitting of nuclei spins by imperfections is
negligible. In the absence of magnetic Zeeman splitting therefore the
nuclear spins in $Ag$ samples will act just as a set of elastic scatterers,
and the temperature dependence of $\tau _{\varphi }$ should obey the
standard theory \cite{AAK82}.

\subsection{Nuclear spins and superconducting order}

The problem of coexistence of the superconducting and magnetic ordering, in
spite of its long history \cite{Ginzb57}, is still among the enigmas of
modern condensed matter physics. Most of the theoretical and experimental
efforts were devoted to studies of the coexistence of electron
ferromagnetism and superconductivity. The possibility of a reduction of $%
H_{c}(T)$ by the nuclear ferromagnetism was outlined by Dyugaev et al. in 
\cite{DVW96} and theoretically studied in more details in \cite{DVW97,KBB97}%
. It was experimentally observed by several groups 
\cite{Rehmann97,Knuuttila01}.

It was recently conjectured \cite{DVW04} that the hyperfine part of the
nuclear-spin-electron interaction may result in the appearance of a
nonuniform superconducting order parameter, the so called
Fulde-Ferrel-Larkin-Ovchinni-kov state (FFLO) \cite{FFLO64}. The FFLO state
was thought originally to take place in superconductors with magnetically
ordered magnetic impurities \cite{FFLO64}. The main difficulty, however, in
the observation of the FFLO caused by magnetic impurity ordering is in the
simultaneous action of the ''electromagnetic'' and ''exchange'' parts of the
magnetic impurities on the superconducting order. In most of the known
superconductors the ''electromagnetic'' part destroys the superconducting
order before the ''exchange'' part modifies the BCS condensate to a
nonuniform FFLO state.

The situation may change drastically in the case of nuclear spin
ferromagnetic ordering. Indeed, the nuclear magnetic moment $\mu _{n}=\hbar
e/M_{i}c,$ is at least three orders of magnitude smaller than the electron
Bohr magneton $\mu _{e}=\hbar e/m_{o}c,$ so that the \textquotedblright
electromagnetic\textquotedblright\ part of the nuclear spin fields is quite
low, compared to that of the magnetic impurities. On the other hand the
\textquotedblright exchange\textquotedblright\ part is strongly dependent on
the nuclear charge $Z$.

\section{Nuclear spins as qubits}

\subsection{Quantum Hall quantum computation}

A growing number of models for electron and nuclear spin based memory cells 
\cite{NucSpMem} and quantum information processing (or Quantum Computing-QC)
has been recently proposed \cite{LdIV98}-\cite{PVW03}. We will concentrate
in what follows on the models based on the manipulation of nuclear spins in
bulk semiconductor \cite{Kane98}, heterostructures \cite{PVK98,MPG01} and
quantum wires and rings \cite{PVW03}.

In \cite{PVK98} we have proposed a quantum computer realization based on
hyperfine interactions between the conduction electrons and nuclear spins
embedded in a two-dimensional electron system in the quantum-Hall effect.
For modifications and improvements of this model see a recent review \cite
{PMV02}.

Maniv et al. \cite{MBVW01} have suggested a following physical process of
preparing a coherent state in QH ferromagnets. Let us assume that at time $%
t=-t_{0}<0$ the filling factor was tuned to a fixed value $\nu =\nu _{0}\neq
1$ and then kept constant until $t=0$ . If $t_{0}\gg T_{2}\left( \nu
_{0}\right) $ then at $t=0$ the nuclear spin system is in the ground state
corresponding to the $2D$ electron system at $\nu =\nu _{0}$. Suppose that
at time $t=0$ the filling factor is quickly switched ( i.e. on a time scale
much shorter than $T_{2}\left( \nu _{0}\right) $ ) back to $\nu =1$ so that
the nuclear spin system is suddenly trapped in its instantaneous
configuration corresponding to $\nu =\nu _{0}\neq 1$ . Thus the nuclear
spins for a long time $t$ ( $\gg $ $T_{2}\left( \nu _{0}\right) $ ) will
find themselves almost frozen in the ground state corresponding to the 2DES
at $\nu =\nu _{0}$, since $T_{2}\left( \nu =1\right) \gg T_{2}\left( \nu
_{0}\right) $.

The main idea behind this scenario stems from the experimental observation
of a dramatic enhancement of the nuclear spin lattice relaxation\ rate $%
1/T_{1}$, and of a sharp decrease of the Knight shift in optically pumped
NMR measurements, as the filling factor is shifted slightly away from $\nu =1
$. The prevailing interpretation of these closely related effects,
associates them with the creation of skyrmions (or antiskyrmions) in the
electron spin distribution as the $2D$ electron system moves away from the
quantum Hall ferromagnetic state at $\nu =1$.

The off-diagonal long range magnetic order \cite{Girvin99} existing in this
state, corresponds to a complex order parameter which is coupled locally
through the hyperfine interaction to the nuclear spins \cite{MBVW01}. It is
thus expected that the proposed manipulation of the nuclear spin system can
be performed in a phase coherent fashion over a spatial region with size of
the order of the skyrmion radius.

The latter effect is still sufficiently weak to enable the survival of a
coherent state of a large number of qubits in the computer memory at $\nu =1$%
. To find an upper bound for the size of such a memory let us consider $N$
independent nuclear spins located at various sites $\mathbf{r}_{j}$ in the
quantum well. A number $n$, stored in the computer memory, corresponds to
the direct product of $N$ pure nuclear spin states $\left\vert
n\right\rangle =$ $\left\vert n_{1}\right\rangle $ $\otimes \left\vert
n_{2}\right\rangle \otimes ...\otimes \left\vert n_{N}\right\rangle $, where 
$\left\vert n_{j}\right\rangle =\sum_{\sigma =\pm 1}\delta _{n_{j},\sigma
}\left\vert j,\sigma \right\rangle $, and $\left\vert j,\sigma \right\rangle 
$ is a nuclear state with spin projection $\sigma $ located at the site $%
\mathbf{r}_{j}$. To be able to start a significant quantum computing
process, however, a coherent superposition of such products, \ i.e. $%
\left\vert \psi \right\rangle =\sum_{n=1}^{N}\alpha _{n}\left\vert
n\right\rangle $ \ (see e.g. \cite{Unruh95}), should be prepared at time $%
t=0 $. \ \ This superposition may be represented more transparently for our
purposes by the direct product of mixed spin up and spin down states, $%
\left\vert \psi \left( t=0\right) \right\rangle =\prod_{j=1}^{N}\otimes
\left( \sum_{\sigma =\pm 1}\alpha _{j,\sigma }\left\vert j,\sigma
\right\rangle \right) $, with the normalization $\sum_{\sigma }\left\vert
\alpha _{j,\sigma }\right\vert ^{2}=1$ . \ The mixing is expected to take
place most efficiently via the flip-flop processes with the electron spins
during the manipulation period when $\nu \neq 1$.

Let us further assume that the initial state $\psi \left( 0\right) $ is a
completely coherent state, so that each of the numbers of length $N$ has an
equal probability. This is a typical state required in carrying out
efficient quantum calculations \cite{Unruh95}. In this state $\left| \alpha
_{j,\sigma }\right| ^{2}=1/2$ for any $j$ and $\sigma .$

It is evident that due to the complete coherence of the initial state $\psi
\left( 0\right) $ the survival probability $P_{\psi }\left( t\right) $

\begin{equation*}
P_{\psi }\left( t\right) =\left[ \frac{1+\func{Re}J\left( t\right) }{2}%
\right] ^{N}\approx \exp \left\{ -\frac{1}{2}N\left[ 1-\func{Re}J\left(
t\right) \right] \right\} \approx e^{-\frac{1}{2}N\Gamma \left( t\right) }
\end{equation*}
depends only on the decoherence factor $J\left( t\right) $. The decay of $%
P_{\psi }\left( t\right) $ therefore follows $\exp [-\frac{1}{2}N\Gamma
\left( t\right) ]$, saturating at $\exp [-\frac{1}{2}N\eta \widetilde{C}%
^{2}] $ for $t$ $\gg $ $\hbar /\varepsilon _{sp}$. Despite the much larger
drop in the level of coherence, \ the time scale over which the coherence
diminishes is the same as in the case of a single qubit.

\subsection{Kane model}

The most popular and intensively developed is the Kane model \cite{Kane98},
where the qubits are the nuclear spin of the $^{31}$P donor placed near to
the surface of \textit{Si}. The nuclear spins are manipulated by electric
gates above the \textit{Si} surface via the atomic electrons. The two-qubit
interaction is provided by the overlapping outer electrons of two
neighboring $^{31}$P donors. The Kane model has stimulated detailed
theoretical and experimental activity.

The strength of the hyperfine interaction is proportional to the probability
density of the electron wavefunction at the nucleus. In semiconductors, the
electron wavefunction extends over large distances through the crystal
lattice. Two nuclear spins can consequently interact with the same electron,
leading to electron-mediated or indirect nuclear spin coupling.

Voltages applied to metallic gates in a semiconductor device may be used to\
control the hyperfine interaction and, correspondingly, the
electron-mediated interaction between the nuclear spins.

A quantum mechanical calculation proceeds by the precise control of three
external parameters:

i) gates above the donors control the strength of the hyperfine interactions
and hence the resonance frequency of the nuclear spins beneath them;

ii) gates between the donors turn on and off electron-mediated coupling
between the nuclear spins;

iii) a globally applied a.c. magnetic field B$_{ac}$ flips nuclear spins at
resonance.

\begin{figure}[tbp]
\begin{center}
\includegraphics[width=4.3094in]{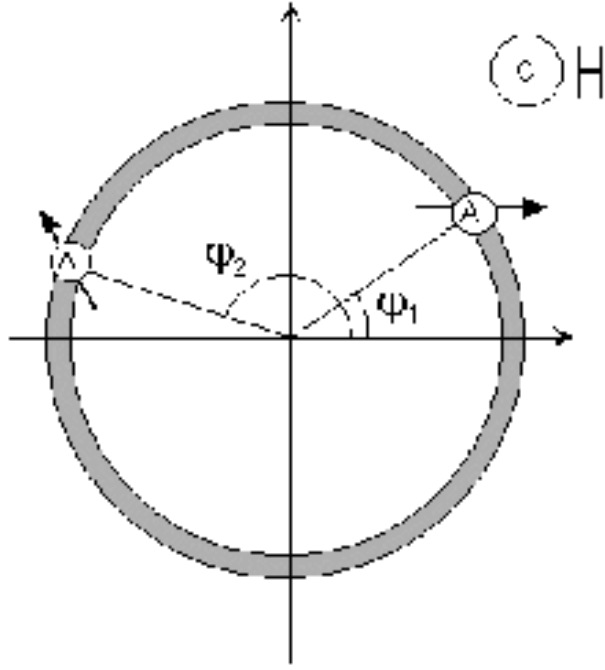} 
\end{center}
\caption{Two-spin-qubits in a nano-ring.}
\end{figure}

Custom adjustment of the coupling of each spin to its neighbors and to B$%
_{ac}$ enables different operations to be performed on each of the spins
simultaneously. Finally, measurements are performed by transferring nuclear
spin polarization to the electrons and determining the electron spin state
by its effect on the orbital wavefunction of the electrons, which can be
probed using capacitance measurements between adjacent gates.

An important requirement for a quantum computer is to isolate the qubits
from any degrees of freedom that may lead to decoherence. If the qubits are
spins on a donor in a semiconductor, nuclear spins in the host are a large
reservoir with which the donor spins can interact. Consequently, the host
should contain only nuclei with spin \textit{I = 0}. This simple requirement
unfortunately eliminates all \textit{III-V} semiconductors as host
candidates, because none of their constituent elements possesses stable 
\textit{I=0} isotopes. Group \textit{IV} semiconductors are composed
primarily of \textit{I =0} isotopes and can in principle be purified to
contain only \textit{I = 0} isotopes.

The only \textit{I = 1/2} shallow (group \textit{V}) donor in Si is $^{31}$%
P. At sufficiently low $^{31}$P concentrations at temperature \textit{T =
1.5 K}, the electron spin relaxation time is thousands of seconds and the $%
^{31}$P nuclear spin relaxation time exceeds 10 hours. It is likely that at
millikelvin temperatures the phonon limited $^{31}$P relaxation time is of
the order of 10$^{18}$ seconds \cite{WS88}, making this system ideal for
quantum computation. Recent calculations of relaxation and decoherence times
for this system are presented in \cite{P31SiHyp}.

\begin{figure}[tbp]
\begin{center}
\includegraphics[width=4.5489in]{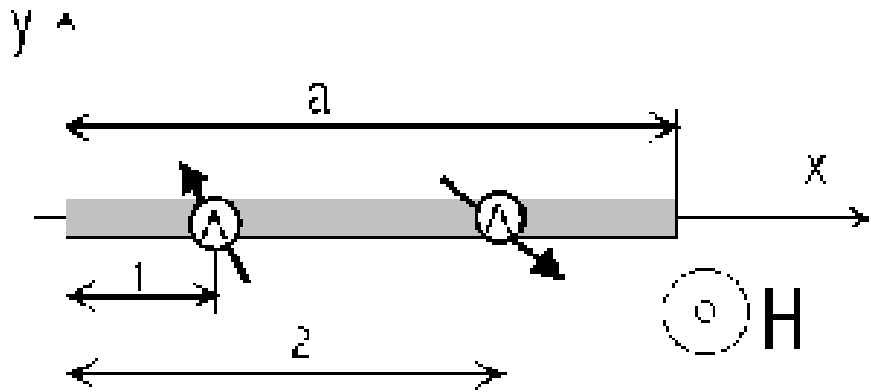} 
\end{center}
\caption{Two spin-qubits in a nano-wire.}
\end{figure}

\subsection{Nuclear-spin qubits in nano-structures}

In \cite{PVW03} a new system is proposed, which consists of nuclear spins
(qubits) embedded into a mesoscopic ring, Fig.~11, or a finite length
quantum wire, Fig.~12. The hyperfine interaction of the electrons in the
system with nuclear spins leads to an effective indirect nuclear spin
interaction

\begin{equation}
E=\left( I_{1,x}I_{2,x}+I_{1,y}I_{2,y}\right) A+I_{1,z}I_{2,z}B\text{,}
\label{inform}
\end{equation}
where $\overrightarrow{I_{i}}$ is magnetic moment of a nucleus, $A$ and $B$
are functions of the system parameters presented in Fig. 13. It is obtained
in \cite{PVW03} that the effective nuclear spins interaction exhibits sharp
maxima, Fig.~13, as function of the magnetic field and nuclear spin
positions which opens the way to manipulate qubits with almost atomic
precision. The selective nuclear spin interaction can be obtained by
changing external parameters of the system, as is shown in Fig.~14.

In \cite{PVW03} the expression for indirect nuclear spin interaction between
nuclear spins in a mesoscopic system, consisting of a finite length quantum
wire or ring with nuclear spins placed in the zero nuclear spin matrix, is
obtained. Interaction between any two qubits, which is necessary for
two-qubit operations, is performed by the electrons in the wire. It is very
sensitive to the system parameters: nuclear spin location, number of
electrons, magnetic field and geometry of the system. Its dependence on the
system parameters is completely different from indirect nuclear spin
interaction in $2D$ and $3D$ metals: by varying the external parameters
(magnetic field and number of electrons) one can control with almost atomic
precision the nuclear spin interaction strength by creating maxima of the
amplitude of electron wave function on some qubits and zero on the other.
The connections to the measuring system, preparation of the initial state
and performing of the one-qubit operations using the NMR could be similar to
the existing experimental suggestions \cite{SSV01}. The decoherence time of
the nuclear spins in mesoscopic systems is expected to be long enough to
perform the quantum computation, since the discrete electron spectrum in
mesoscopic systems imposes restriction on the flip-flop processes and the
nuclear spin relaxation time at law temperatures is expected to have an
activation behavior.

\begin{figure}[tbp]
\begin{center}
\includegraphics[width=4.0698in]{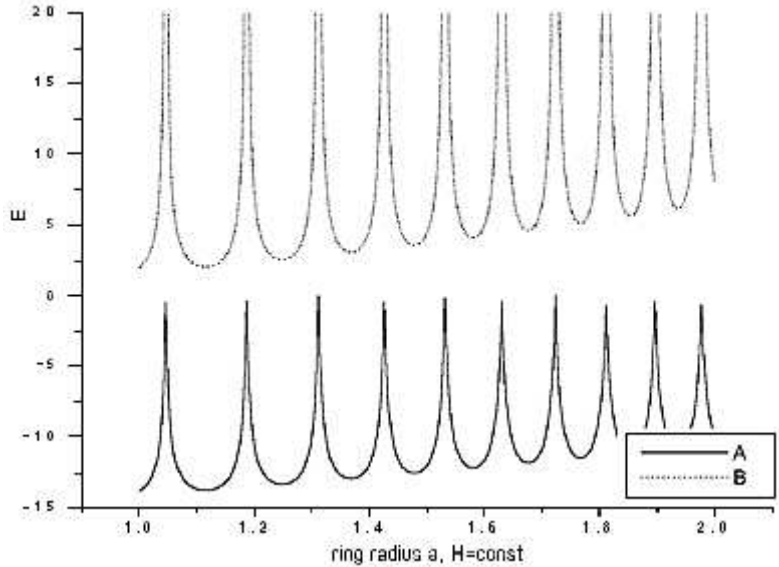} 
\end{center}
\caption{Interaction energy coefficients
A and B, Eq. (\ref{inform}), as function of the magnetic flux through the
ring.}
\end{figure}

\begin{figure}[tbp]
\begin{center}
\includegraphics[width=4.6259in]{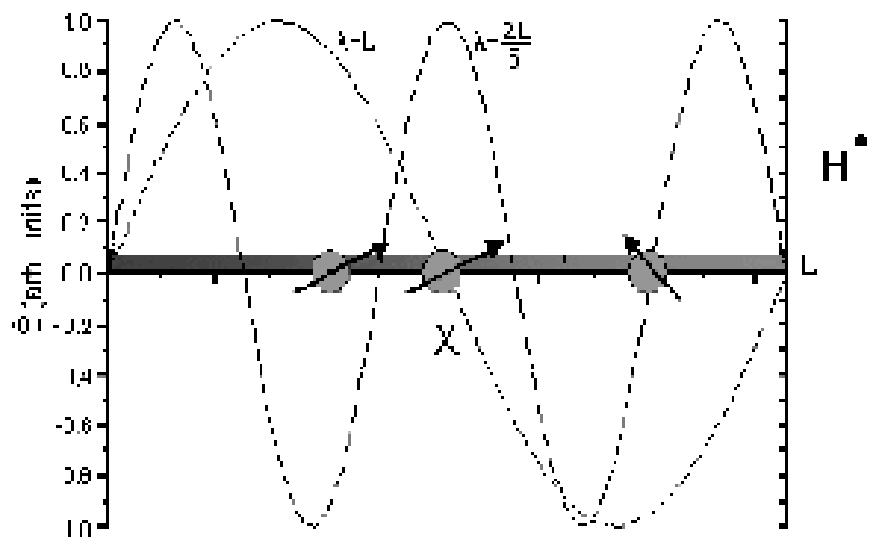} 
\end{center}
\caption{Indirect interaction via standing
wave like conduction electron wave function between spin-qubits, embedded in
a nano-wire of a finite length.}
\end{figure}

\section{Experiments}

The measurement of the nuclear spin-relaxation in heterojunctions is a
challenging experimental problem, since the direct detection of the NMR
signals in solids requires usually $10^{17}-10^{20}$ nuclei. The number of
nuclear spins interacting with the two-dimensional electrons is however,
much smaller: $10^{12}-10^{15}$.

The first successful measurements of the magnetic field dependence of $%
T_{1}^{-1}$ under QHE conditions were performed in a series of elegant
experiments by the K. von Klitzing group, \cite{DKSWP88}. Combining
electron-spin resonance (ESR) and resistivity measurement techniques they
have observed the shifting of the ESR resonance frequency by the hyperfine
field of nonequilibrium nuclear spin population, which is the well known
Overhauser shift \cite{AbragBk61}. In this experiment the $2D$ electron
Zeeman splitting is tuned to the pumping frequency. The angular momentum
gained by a 2DEG electron, excited to the upper Zeeman branch, is then
transferred to the nuclear spins, thus creating a nonequilibrium nuclear
spin population.

These measurements show a close similarity between the magnetic field
dependence of the nuclear spin-relaxation rate and the magnetoresistance in
quantum Hall effect, as it was suggested theoretically in \cite{VM88}, thus
demonstrating clearly the importance of the coupling of nuclear spins to the
conduction electron spins in the nuclear relaxation processes in these
systems.

Various experimental techniques were used since and in what follows we will
describe shortly the main developments and achievements in experimental
studies of the hyperfine coupling between the nuclear spins and the
electrons in QHE, mesoscopic and superconducting systems.

Another way of measuring the nuclear spin relaxation and diffusion in a
heterojunction under strong magnetic field by transport techniques
(spin-diode) was demonstrated by Kane et al. \cite{KPW92}. They have
reported measurements performed on ``spin diodes`` : junctions between two
coplanar 2DEG's in which $\nu <$l on one side and $\nu >$ 1 on the other.
The Fermi level E$_{F}$ crosses between spin levels at the junction. In such
a device the 2DEG is highly conducting except in a narrow region (with a
width of the order of several hundred angstroms) where $\nu =1.$

Wald et al. \cite{Wald94} presented the experimental evidence for the
effects of nuclear spin diffusion and the electron-nuclear Zeeman
interaction on interedge state scattering. Polarization of nuclear spins by
dc current has proven to be a rich source for new, not yet understood
completely, phenomena. This is the case, for example of anomalous spikes in
resistivity around certain fractional filling factors, observed by
Kronmuller et al. \cite{Kronm99} and in resistively detected NMR \ in QHE
regime reported by Desrat et al. \cite{Desrat02}. \ Influence of
nonequilibrium nuclear spin polarization on Hall conductivity and
magnetoresistance was observed and studied in detail by Gauss et al.~\cite
{GJ96}.

In 1994 Barrett et al. \cite{Barrett94} observed, for the first time, a
sharp NMR signal in multi-quantum wells, using the Lampel \cite{Lampel68}
technique of polarizing the nuclear spins by optical pumping of interband
transitions with near-infrared laser light \textit{(OPNMR)}. Polarization of
nuclei results in a significant enhancement of the NMR sensitivity, since
the resonance in a two-level system results in equalizing their population.
The difference in the population is, obviously, maximal when the spins are
completely polarized.

Detailed studies of the Knight shift data suggested \cite{Barrett94} that
the usually accepted picture of electron spins, aligned parallel to the
external field, should be modified to include the possibility of
topologically nontrivial nuclear spin orientations, the Skyrmions. Optical
polarization of nuclear spins was used also as a tool for reducing the
Zeeman splitting of $2D$ electrons by Kukushkin et al. \cite{KKE99}. This
resulted in a noticeable enhancement of the skyrmionic excitations. Similar
results are reported by Vitkalov et al. \cite{Vitkalov00}.

Very recently, a modern ultra-sensitive NMR spin-echo technique was employed
to study the physics of QHE \cite{Melinte99} in GaAs/AlGaAS multi-quantum
well heterostructures. The spin polarization of 2DES in the quantum limit
was investigated and the experimental data support the noninteracting
Composite Fermion model in the vicinity of the filling factor $\nu =1/2$ . \
Using the same technique the polarization of 2DES near $\nu =2/3$ was
investigated \cite{Freitag01}. It was shown there that a quantum phase
transition from a partially polarized to a fully polarized state can be
driven by increasing the ratio between the Zeeman and Coulomb energies.

An amazing phenomenon, following from the hyperfine coupling between the
electron and the nuclear spins, is the giant enhancement of the low
temperature heat capacity of GaAs quantum wells near the filling factor $\nu
=1$, discovered in 1996 by Bayot et al. \cite{BGMSS96}. \ As other
thermodynamic properties, it experiences quantum oscillations, following
from the oscillatory density of states $D\left( E_{F}\right) $ at the Fermi
level.

At about $T=25mK$, and in clean samples, Bayot et al. \cite{BGMSS96} have
observed anomalous deviations from the free electron model, in the specific
heat value (up to four orders of magnitude) for the filling factor in the
range $0.5\leq \nu $ $\leq 1.5$ . Their explanation is that in this interval
of parameters, the electron system couples strongly to nuclear spin system
with a concentration of several orders of magnitude larger than the electron
one.

This raises a question about the origin of the strong coupling between the
electron and the nuclear spins in the interval $0.5\leq \nu $ $\leq 1.5$ .
The guess is the skyrmions, since they are predicted to appear just in the
same interval of the filling factor. Additional support for this mechanism
is in the results of \cite{Melinte99}, where the disappearance of the
nuclear spin contribution to the heat capacity was reported, as the ratio
between the Zeeman and Coulomb energies exceeds a certain critical value.
The Zeeman splitting of electrons was modified in these experiments by
tilting the magnetic field.

A new very promising technique for measuring spatially varying nuclear spin
polarization within a GaAs sample is reported in \cite{Thurber02}. In the
force detected NMR (FDNMR) the sample is mounted on a microcantilever in an
applied magnetic field. A nearby magnetic particle creates a gradient of
magnetic field which exerts a force on the magnetized sample and triggers
the cantilever oscillations. FDNMR is capable to perform the magnetic
resonance imaging of the sample with a very high accuracy. \ This method can
be useful in defining the spatial distribution of nuclear spin polarization
in non homogeneous samples. This information may be crucial for
understanding different peculiarities of data obtained by previously
described methods.

A new world of the low-temperature physics of the hyperfine interactions in
superconducting metals opens in the $\mu K$ region, where the nuclear spins
start ordering, thus reducing the critical magnetic field of superconductors
as has been recently discovered by the Pobell group \cite{Rehmann97}. They
have studied the magnetic critical field $H_{c}(T)$ of a metallic compound $%
\ AuIn_{2}$ where the superconductivity sets up at $T_{ce}=0.207K.$ \ They
have observed, in $AuIn_{2}$, the nuclear spin ferromagnetic ordering at $%
T_{cn}=35\mu K.$ It was observed in these experiments that the magnetic
critical field $H_{c0}=14.5$ G is lowered by almost a factor of two at $%
T<T_{cn}.$

\newpage

\section*{Acknowledgements}

In writing this review the author benefited enormously from the colleagues
and co-workers: P.~Averbuch, M.~Azbel, S.E.~Barret, C.~Berthier,
Yu.~Bychkov, A.~Dyugaev, E.~Ehrenfreund, V.I.~Fal'ko, V.~Fleurov,
L.P.~Gor'kov, V.L.~Gurevich, M.~Horvati\'{c}, S.V.~Iordanskii,
A.G.M.~Jansen, K.~von~Klitzing and his group, K.~Kikoin, Ju.H.~Kim,
B.I.~Lembrikov, D.~Loss, T.~Maniv, S.~Meshkov, D.~Mozyrsky, Yu.~Ovchinnikov,
Yu.~Pershin, V.~Privman, F.~Pobell, M.~Potemski, late A.~Rozhavsky, A.~Shik,
I.~Shlimak, B.~Spivak, P.C.E.~Stamp, P.~Wyder, A.~Zyuzin, whose knowledge,
intuition and professional skills were crucial in going through the jungle
of hyperfine interactions in such exotic systems, as superconducting,
quantum Hall, mesoscopic and nano-sytems.

My gratitude to Alexander Kaplunovsky for the numerical computations and to
Pinchas Malits for mathematical support.

This work is supported by grants from: ISF, the Israeli Science Foundation
and INTAS: 2001-0791.

\end{document}